\documentclass[twocolumn]{aastex631}

\usepackage[utf8]{inputenc}
\DeclareUnicodeCharacter{2212}{-}
\usepackage{graphicx}
\graphicspath{{./}{./figures/}{./tex/figures/}}
\usepackage{amsmath}
\usepackage{mathtools}
\usepackage{afterpage}
\usepackage{enumitem}
\usepackage{xcolor}
\setenumerate{itemsep=0mm}

\newcommand{\norm}{\mathcal{N}}
\newcommand{\gaia}{\emph{Gaia}\xspace}

\newcommand{\delve}{DELVE\xspace}
\newcommand{\SSSSS}{\ensuremath{S^5}\xspace}
\newcommand{\Jet}{\ensuremath{\mathrm{Jet}}\xspace}

\usepackage{soul} 
\usepackage{amsmath}
\usepackage{amssymb}
\usepackage{xspace}
\usepackage{xifthen}
\usepackage{eso-pic}

\newcommand{\Gaia}{{\it Gaia}\xspace}

\definecolor{forestgreen}{HTML}{228B22}
\definecolor{urlblue}{HTML}{000000}

\newcommand{\CHECK}[1]{{\textcolor{black}{#1}}}
\newcommand{\response}[1]{#1}

\mathchardef\mhyphen="2D

\newcommand{\roughly}{\ensuremath{ {\sim}\,} }

\newlength{\dhatheight}

\newcommand{\code}[1]{\texttt{#1}\xspace}

\newcommand{\unit}[1]{\ensuremath{\mathrm{\,#1}}\xspace}
\newcommand{\yr}{\unit{yr}}
\newcommand{\Gyr}{\unit{Gyr}}
\newcommand{\Myr}{\unit{Myr}}

\newcommand{\degree}{\ensuremath{{}^{\circ}}\xspace}

\newcommand{\mas}{\unit{mas}}
\newcommand{\amin}{\unit{arcmin}}
\newcommand{\asec}{\unit{arcsec}}

\newcommand{\km}{\unit{km}}

\newcommand{\pc}{\unit{pc}}
\newcommand{\kpc}{\unit{kpc}}
\newcommand{\second}{\unit{s}}

\newcommand{\Msun}{\unit{M_\odot}}

\newcommand{\magn}{\unit{mag}}

\newcommand{\e}{\unit{e^{-}}}

\newcommand{\secref}[1]{Section~\ref{sec:#1}}

\newcommand{\tabref}[1]{Table~\ref{tab:#1}}

\newcommand{\figref}[1]{Figure~\ref{fig:#1}}

\newcommand{\bandvar}[2][]{%
  \ifthenelse{\isempty{#1}}{\var{#2}}{\var{#2\_#1}}%
}

\newcommand{\LCDM}{\ensuremath{\rm \Lambda CDM}\xspace}

\newcommand{\SExtractor}{\code{SExtractor}}

\newcommand{\PSFex}{\code{PSFex}}

\newcommand{\HEALPix}{\code{HEALPix}}
\newcommand{\healpix}{\HEALPix}

\newcommand{\emcee}{\code{emcee}}
\newcommand{\ugali}{\code{ugali}}

\newcommand{\var}[1]{\ensuremath{\texttt{\MakeUppercase{#1}}}\xspace}
\newcommand{\nside}{\code{nside}}

\newcommand{\given}{\ensuremath{ \,|\, }\xspace}

\providecommand\physrep{\ref@jnl{Phys.~Rep.}}%
\providecommand\apjs{\ref@jnl{ApJS}}%
\providecommand{\jcap}{\ref@jnl{JCAP}}%

\newcommand{\initaltrackE}{-0.00002}
\newcommand{\initaltrackD}{0.00006}
\newcommand{\initaltrackC}{-0.00138}
\newcommand{\initaltrackB}{0.01475}
\newcommand{\initaltrackA}{0.07247}

\newcommand{\dmtrackA}{17.45}
\newcommand{\dmtrackB}{-0.014}

\newcommand{\fehjet}{$-1.57$}

\newcommand{\pmMuAStr}{-0.933}
\newcommand{\pmMuBStr}{-0.023}
\newcommand{\dpmMuAStr}{-0.109}
\newcommand{\dpmMuBStr}{0.120}
\newcommand{\fStr}{0.086}

\newcommand{\pmMuAStrUP}{0.018}
\newcommand{\pmMuBStrUP}{0.016}
\newcommand{\dpmMuAStrUP}{0.030}
\newcommand{\dpmMuBStrUP}{0.021}
\newcommand{\fStrUP}{0.011}

\newcommand{\highnumRGBcandpm}{89}
\newcommand{\mednumRGBcandpm}{97}
\newcommand{\highnumBHBcandpm}{28}
\newcommand{\mednumBHBcandpm}{28}

\newcommand{\rperi}{12.7}
\newcommand{\rperierr}{0.9}
\newcommand{\rapo}{38}
\newcommand{\rapoerr}{1}
\newcommand{\lz}{1.9}
\newcommand{\lzerr}{0.1}
\newcommand{\energy}{-0.103}
\newcommand{\energyerr}{0.001}

\definecolor{newblue}{cmyk}{1,0.7,0,0}
\definecolor{mypink}{cmyk}{0, 0.7808, 0.4429, 0.1412}

\shorttitle{The Jet Stream in Delve}

\shortauthors{DELVE Collaboration}

\begin{document}

\reportnum{\footnotesize FERMILAB-PUB-21-205-AE-LDRD}

\title{DELVE-ing into the Jet: a thin stellar stream on a retrograde orbit at 30 kpc}


\author[0000-0001-6957-1627]{P.~S.~Ferguson}
\affiliation{George P. and Cynthia Woods Mitchell Institute for Fundamental Physics and Astronomy, Texas A\&M University, College Station, TX 77843, USA}
\affiliation{Department of Physics and Astronomy, Texas A\&M University, College Station, TX 77843, USA}
\author[0000-0003-2497-091X]{N.~Shipp}
\affiliation{Department of Astronomy and Astrophysics, University of Chicago, Chicago IL 60637, USA}
\affiliation{Kavli Institute for Cosmological Physics, University of Chicago, Chicago, IL 60637, USA}
\author[0000-0001-8251-933X]{A.~Drlica-Wagner}
\affiliation{Fermi National Accelerator Laboratory, P.O.\ Box 500, Batavia, IL 60510, USA}
\affiliation{Kavli Institute for Cosmological Physics, University of Chicago, Chicago, IL 60637, USA}
\affiliation{Department of Astronomy and Astrophysics, University of Chicago, Chicago IL 60637, USA}
\author[0000-0002-9110-6163]{T.~S.~Li}
\affiliation{Observatories of the Carnegie Institution for Science, 813 Santa Barbara St., Pasadena, CA 91101, USA}
\affiliation{Department of Astrophysical Sciences, Princeton University, Princeton, NJ 08544, USA}
\affiliation{NHFP Einstein Fellow}
\author[0000-0003-1697-7062]{W.~Cerny}
\affiliation{Kavli Institute for Cosmological Physics, University of Chicago, Chicago, IL 60637, USA}
\affiliation{Department of Astronomy and Astrophysics, University of Chicago, Chicago IL 60637, USA}
\author[0000-0001-6584-6144]{K.~Tavangar}
\affiliation{Kavli Institute for Cosmological Physics, University of Chicago, Chicago, IL 60637, USA}
\affiliation{Department of Astronomy and Astrophysics, University of Chicago, Chicago IL 60637, USA}
\author[0000-0002-6021-8760]{A.~B.~Pace}
\affiliation{McWilliams Center for Cosmology, Carnegie Mellon University, 5000 Forbes Ave, Pittsburgh, PA 15213, USA}
\author[0000-0003-0710-9474]{J.~L.~Marshall}
\affiliation{George P. and Cynthia Woods Mitchell Institute for Fundamental Physics and Astronomy, Texas A\&M University, College Station, TX 77843, USA}
\affiliation{Department of Physics and Astronomy, Texas A\&M University, College Station, TX 77843, USA}
\author[0000-0002-7134-8296]{A.~H.~Riley}
\affiliation{George P. and Cynthia Woods Mitchell Institute for Fundamental Physics and Astronomy, Texas A\&M University, College Station, TX 77843, USA}
\affiliation{Department of Physics and Astronomy, Texas A\&M University, College Station, TX 77843, USA}
\author[0000-0002-6904-359X]{M.~Adam\'ow}
\affiliation{National Center for Supercomputing Applications, University of Illinois, 1205 West Clark St., Urbana, IL 61801, USA}
\affiliation{Center for Astrophysical Surveys, National Center for Supercomputing Applications, Urbana, IL, 61801, USA}
\author[0000-0002-3936-9628]{J.~L.~Carlin}
\affiliation{Rubin Observatory/AURA, 950 North Cherry Avenue, Tucson, AZ, 85719, USA}
\author[0000-0003-1680-1884]{Y.~Choi}
\affiliation{Space Telescope Science Institute, 3700 San Martin Drive, Baltimore, MD 21218, USA}
\author[0000-0002-8448-5505]{D.~Erkal}
\affiliation{Department of Physics, University of Surrey, Guildford GU2 7XH, UK}
\author[0000-0001-5160-4486]{D.~J.~James}
\affiliation{ASTRAVEO, LLC, PO Box 1668, Gloucester, MA 01931}
\author[0000-0003-2644-135X]{Sergey~E.~Koposov}
\affiliation{Institute for Astronomy, University of Edinburgh, Royal Observatory, Blackford Hill, Edinburgh EH9 3HJ, UK}
\affiliation{Institute of Astronomy, University of Cambridge, Madingley Road, Cambridge CB3 0HA, UK}
\affiliation{Kavli Institute for Cosmology, University of Cambridge, Madingley Road, Cambridge CB3 0HA, UK}
\author[0000-0003-2511-0946]{N.~Kuropatkin}
\affiliation{Fermi National Accelerator Laboratory, P.O.\ Box 500, Batavia, IL 60510, USA}
\author[0000-0002-9144-7726]{C.~E.~Mart\'inez-V\'azquez}
\affiliation{Cerro Tololo Inter-American Observatory, NSF's National Optical-Infrared Astronomy Research Laboratory,\\ Casilla 603, La Serena, Chile}
\author[0000-0003-3519-4004]{S.~Mau}
\affiliation{Department of Physics, Stanford University, 382 Via Pueblo Mall, Stanford, CA 94305, USA}
\affiliation{Kavli Institute for Particle Astrophysics \& Cosmology, P.O.\ Box 2450, Stanford University, Stanford, CA 94305, USA}
\author[0000-0001-9649-4815]{B.~Mutlu-Pakdil}
\affiliation{Kavli Institute for Cosmological Physics, University of Chicago, Chicago, IL 60637, USA}
\affiliation{Department of Astronomy and Astrophysics, University of Chicago, Chicago IL 60637, USA}
\author[0000-0002-7134-8296]{K.~A.~G.~Olsen}
\affiliation{NSF's National Optical-Infrared Astronomy Research Laboratory, 950 N. Cherry Ave., Tucson, AZ 85719, USA}
\author[0000-0002-1594-1466]{J.~D.~Sakowska}
\affiliation{Department of Physics, University of Surrey, Guildford GU2 7XH, UK}
\author[0000-0003-1479-3059]{G.~S.~Stringfellow}
\affiliation{Center for Astrophysics and Space Astronomy, University of Colorado, 389 UCB, Boulder, CO 80309-0389, USA}
\author[0000-0002-9541-2678]{B.~Yanny}
\affiliation{Fermi National Accelerator Laboratory, P.O.\ Box 500, Batavia, IL 60510, USA}

\collaboration{23}{(DELVE Collaboration)}

\correspondingauthor{Peter Ferguson}
\email{peter.ferguson@wisc.edu}

\begin{abstract}
We perform a detailed photometric and astrometric analysis of stars in the Jet stream using data from the first data release of the DECam Local Volume Exploration Survey (DELVE) DR1 and \Gaia EDR3. 
We discover that the stream extends over $\roughly 29 \degree$ on the sky (increasing the known length by $18 \degree$), which is comparable to the kinematically cold Phoenix, ATLAS, and GD-1 streams. 
Using blue horizontal branch stars, we resolve a distance gradient along the Jet stream of 0.2 kpc/deg, with distances ranging from $D_\odot \roughly 27-34$ kpc.
We use natural splines to simultaneously fit the stream track, width, and intensity to quantitatively characterize density variations in the Jet stream, including a large gap, and identify substructure off the main track of the stream.  
Furthermore, we report the first measurement of the proper motion of the Jet stream and find that it is well-aligned with the stream track suggesting the stream has likely not been significantly perturbed perpendicular to the line of sight.
Finally, we fit the stream with a dynamical model and find that the stream is on a retrograde orbit, and is well fit by a gravitational potential including the Milky Way and Large Magellanic Cloud.
These results indicate the Jet stream is an excellent candidate for future studies with deeper photometry, astrometry, and spectroscopy to study the potential of the Milky Way and probe perturbations from baryonic and dark matter substructure. 
\end{abstract}



\section{Introduction}
\label{sec:intro}

Stellar streams form through the tidal disruption of dwarf galaxies and globular clusters as they accrete onto a larger host galaxy \citep[e.g.,][]{Newberg:2016}. 
The formation of stellar streams is an expected feature of hierarchical models of galaxy formation where large galaxies grow through mergers of smaller systems \citep{Lynden-Bell:1995, Johnston:2001}. 
Due to their formation mechanism, transient nature, and dynamical fragility, stellar streams provide a direct and powerful probe of the gravitational field in galactic halos at both large and small scales \citep[e.g.,][]{Johnston:1999,Johnston:2002,Ibata:2002}. 
Within our Milky Way in particular, stellar streams have been proposed as sensitive probes of the large- and small-scale distributions of baryonic and dark matter within the Galactic halo~\citep{Johnston:2001,Carlberg:2013,Erkal:2016,Bonaca:2018,Banik:2019}.

Milky Way stellar streams form when stars are unbound from the progenitor at the Lagrange points between the progenitor stellar system and the Milky Way. 
Stars that are unbound from the inner Lagrange point have lower energy and thus shorter orbital periods than the progenitor whereas those at the outer Lagrange point have higher energy and shorter orbital periods. 
Thus, as the progenitor is disrupted, leading and trailing streams of stars will form roughly tracing the orbit of the progenitor within the Milky Way potential \citep{Sanders:2013}. 
The width of a stellar stream is proportional to the velocity dispersion of its progenitor \citep{Johnston:2001,Erkal:2019}, implying that stellar streams formed through the disruption of globular clusters are narrow ($\roughly 100 \pc$) and dynamically cold, while streams originating from dwarf galaxies are broader ($>500\pc$) and dynamically hot. 
The population of cold stellar streams with small internal velocity dispersions provides a sensitive probe of the gravitational field far from the Milky Way disk.

In a smooth gravitational potential, stellar streams form as coherent structures spanning tens of degrees on the sky \citep[e.g.,][]{Newberg:2016}. 
Long stellar streams can be used to trace the local gravitational field over tens of kpc \citep{Bovy:2014}. 
In conjunction with orbit modeling and simulations, streams can constrain the total mass enclosed inside their orbits \citep[e.g.,][]{Gibbons:2014,Bowden:2015MNRAS.449.1391B, Bovy:2016ApJ...833...31B, Bonaca:2018, Malhan:2019}, and the shapes and radial profiles of the gravitational field \citep[e.g.,][]{Law:2010ApJ...714..229L, koposov:2010ApJ...712..260K, Bowden:2015MNRAS.449.1391B, Malhan:2019}.  
\citet{Bonaca:2018} find that a dozen cold stellar streams with full 6D kinematic measurements should contain enough information to constrain the mass and shape of a simple Milky Way potential with $\sim 1 \%$ precision. 
Additionally, perturbations from large structures can induce a misalignment between the orbit and track of a stream, which can be used to constrain the mass of the perturbing object \citep{Erkal:2018,shipp:2019ApJ...885....3S}.  
For example, \citet{Erkal:2019} and \citet{Vasiliev:2021} used the Orphan and Sagittarius streams, respectively, to simultaneously measure the mass of the Milky Way and Large Magellanic Cloud (LMC).

Stellar streams can also probe the clustering and distribution of dark matter at small scales.
The dark energy plus cold dark matter (\LCDM) model predicts that dark matter should clump into gravitationally bound halos on scales that are much smaller than the smallest galaxies \citep{Green:2004,Diemand:2005,Wang:2020}.
Dark matter subhalos that pass close to stellar streams may gravitationally perturb the stream by altering the stream track and inducing small-scale density fluctuations.  
Discrete gaps in stellar streams, such as those found in the Pal 5 and GD-1 streams discovered from data collected by the Sloan Digital Sky Survey \citep[][]{Odenkirchen:2001,Grillmair:2006}, can probe the population of compact subhalos with $10^6 \Msun < M < 10^8 \Msun$ that contain no luminous matter \citep[e.g.,][]{Erkal:2016,Bonaca:2019}. 
Additionally, the power spectrum of density fluctuations along a cold stream can place limits on the number of dark subhalos \citep[e.g.,][]{Banik:2019} and the mass of warm dark matter candidates \citep[e.g., sterile neutrinos;][]{Dodelson:1994,Shi:1999}. However, baryonic structures such as giant molecular clouds \citep{Amorisco:2016} or Milky Way substructure such as the disk and bar \citep{Erkal2017:pal5MNRAS.470...60E, Pearson:2017,Banik:2019} can induce perturbations that mimic the observational signature of dark matter subhalos. 
It is thus crucial to characterize cold streams at large Galactocentric radii where they are less likely to be affected by baryonic structures \citep[e.g.,][]{Li:2020}.

Despite the importance of Milky Way stellar streams as probes of galaxy formation in a cosmological context, they remain difficult to detect due to their low surface brightness (fainter than $28.5 \; \mathrm{mag/arcsec}^2$) and large spatial extent across the sky ($\gtrsim 10 \degree$).
The phase space signature of streams at large Galactocentric distances is often difficult to detect from space-based observatories (e.g., \Gaia). Stars in these streams are either too faint to have well-measured proper motions or their proper motions overlap with the locus of faint foreground stars at small distances \citep{Ibata:2020}. Distant streams have only recently been detected thanks to deep, wide-area imaging by ground-based digital sky surveys \citep[e.g. SDSS, Pan-STARRS1, and DES,][]{Belokurov:2006, Bernard:2016, Shipp:2018}.

The Jet stream is one such dynamically cold stellar stream that was discovered by \citet{JethwaJetstream:2018} (hereafter referred to as J18) in the Search for the Leading Arm of Magellanic Satellites (SLAMS) survey. This stream was found to have a width of 0.18\degree and a length of 11\degree (truncated on one end by the survey footprint). They found the stellar population of Jet to be well-described by an old (12.1~Gyr), metal-poor ($\text{[Fe/H]}=$\fehjet) isochrone. Fits to the main sequence turn-off (MSTO) and the distribution of blue horizontal branch (BHB) stars in the central portion of the stream place its heliocentric distance at $\sim$29~kpc. 
At this distance the physical width of the stream corresponds to $\sim$ 90~pc, placing the stream firmly in the dynamically cold category. 
This narrow width also suggests the progenitor of Jet was likely a globular cluster, although no progenitor was found by J18.

We further investigate the Jet stream using data from the DECam Local Volume Exploration Survey (DELVE) Data Release 1 (DR1) \citep{Drlica-Wagner:2021}.  
This catalog covers over $\sim$ 4,000 $\text{deg}^2$ in four photometric bands ($g,r,i,z$) and over $\sim$ 5,000 $\text{deg}^2$ in each band independently. 
The sensitivity of DELVE has been demonstrated by the discovery of a Milky Way satellite galaxy candidate with $M_V = -5.5$ at a distance of $\roughly 116 \kpc$ \citep[Centaurus I;][]{mau:2020} and two faint star cluster candidates \citep[DELVE 1, DELVE 2;][]{mau:2020,cerny:2020}. 
DELVE DR1 contiguously and homogeneously covers a large region including and extending the SLAMS survey footprint. 
Thus, the DELVE data are ideal to further characterize the Jet stream.

To dynamically model the stream and extract local properties of the gravitational field, additional phase space information is needed.
With full 3D kinematic information, the Jet stream can become an even better tool for measuring the properties of the Milky Way, thereby allowing us to probe its interaction history.
A combination of proper motion and radial velocity measurements are required to obtain the full 6D phase space information of the Jet stream.
The high-precision astrometric survey \Gaia has revolutionized this field and allowed for measurements of the proper motion of faint stream stars for the first time. 
The early third data release from \Gaia \citep[\Gaia EDR3;][]{Gaia:2016A&A...595A...1G,GaiaeDR3:2020arXiv201201533G} provides proper motion measurements for more than 1.4 billion stars down to a magnitude of $G\sim21$. \Gaia has previously been used to characterize the proper motions of many stellar streams \cite[e.g.][]{shipp:2019ApJ...885....3S,Price-whelan:2018ApJ...863L..20P,koposov2019:orphanMNRAS.485.4726K}
and discover tens of candidate stellar streams \cite[e.g.][]{Malhan:2018a,Malhan:2018b,Ibata:2020}.  
In this paper we use astrometric measurements from \Gaia and photometry from \delve to measure the proper motion of the Jet stream for the first time and quantitatively characterize its shape. 
These measurements, along with future spectroscopic observations, will allow for a full characterization and dynamical modeling of this stream. 

This paper is organized as follows. In \secref{data} we briefly present the DELVE DR1 dataset. We then describe our analysis of the Jet stream in \secref{methods}, initially using the DELVE DR1 dataset to characterize the stream track over an extended region of the sky.
Next, we measure a distance gradient along the stream using blue horizontal branch stars, and use this distance gradient to optimize our matched-filter. 
Then, we model the observations to quantitatively characterize the structure of the stream. 
To further characterize the stream, we use the DELVE DR1 photometry along with proper motion measurements from \Gaia EDR3 to measure the proper motion of the Jet stream for the first time. 
In Section \ref{sec:modeling} we fit the stream with a dynamical model to determine the best-fit orbital parameters and to determine whether the stream is likely to have been significantly perturbed by large substructure such as the Milky Way bar or presence of the LMC.   
We discuss our results in \secref{Discussion} and conclude in \secref{conclusions}.

\section{DELVE DR1 Data}
\label{sec:data}
\begin{figure}[ht]
    \centering
    \includegraphics[width=0.98\columnwidth ]{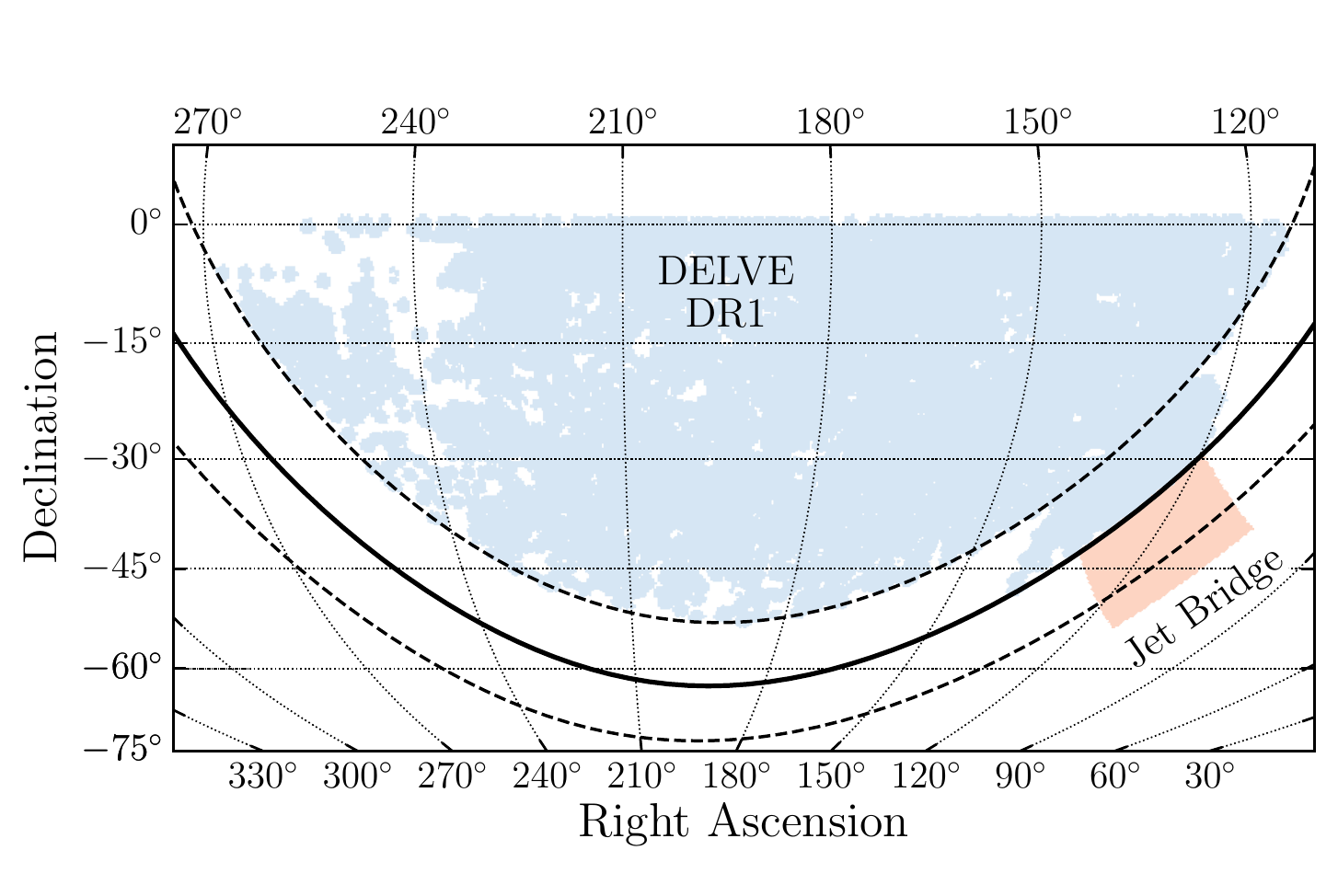}
    \caption{The DELVE DR1 region where our search was performed is shown in light blue. The additional region (the Jet Bridge) that was searched is shown as a light orange patch. The solid black line indicates the plane of the Milky Way ($b=0\degree$) and the two dashed lines indicate $b=\pm10\degree$. 
    }
    \label{fig:search_region}
\end{figure}

DELVE seeks to provide contiguous and homogeneous coverage of the high-Galactic-latitude ($|b| >10\degree$) southern sky ($\delta < 0\degree$) in the $g,r,i,z$ bands \citep{Drlica-Wagner:2021}. This is done by assembling all existing archival DECam data and specifically observing regions of the sky that have not been previously observed by other community programs. These data are consistently processed with the same data management pipeline to create a uniform dataset \citep{Morganson:2018}. The DELVE DR1 footprint consists of the region bounded by $\delta < 0\degree$ and $b >10\degree$ with an additional extension to $b = 0\degree$ in the region of $120\degree < \alpha < 140\degree$ to search for extensions of the Jet stream. This footprint is shown in Figure \ref{fig:search_region} as a light blue shading. Additionally, we searched the Jet Bridge region below the Galactic plane (light orange shading). 
However, no evidence of a continuation of the Jet stream was found in this region. 

The DELVE DR1 dataset consists of $\sim$30,000 DECam exposures, with exposure times between $30\second < {\rm t_{exp}} < 350\second$.
Additionally, the following quality cuts were applied to individual exposures: a minimum cut on the effective exposure time scale factor $\mathrm{t_{eff}} > 0.3$ \citep{Neilsen:2015} and a good astrometric solution relative to \Gaia DR2 (For each exposure $>250$ astrometric matches, $\chi^2_{\rm astrom} < 500$, where a match has $\Delta(\alpha) < 150\mas$, and $\Delta(\delta) < 150\mas$). 
All exposures were processed with the DES Data Management (DESDM) pipeline \citep{Morganson:2018}, enabling sub-percent-level photometric accuracy by calibrating based on seasonally averaged bias and flat images and performing full-exposure sky background subtraction \citep{Bernstein:2018}.  
Automatic source detection and photometric measurement is performed on each exposure using \SExtractor and \PSFex \citep{Bertin:1996, Bertin:2011}.
Astrometric calibration was performed against \Gaia DR2 using \code{SCAMP} \citep{Bertin:2006}.
Photometric zeropoints for each CCD were derived by performing a $1\asec$ match between the DELVE \SExtractor catalogs and the ATLAS Refcat2 catalog \citep{Tonry:2018},
and using transformation equations derived by comparing stars in ATLAS Refcat2 to calibrated stars from DES DR1 to convert the ATLAS Refcat2 measurements into the DECam \emph{griz}-bandpass \citep{Drlica-Wagner:2021}.
The zeropoints derived from this processing were found to agree with the DES DR1 zeropoints with a scatter of $\lesssim 0.01 \magn$.
Dust extinction corrections were applied using extinction maps from \citet{Schlegel:1998} assuming $R_V=3.1$ and a set of $R_\lambda$ coefficients derived by DES \citep{DES:2018} including a normalization adjustment from \citet{Schlafly:2011}. 
Hereafter, all quoted magnitudes have been corrected for interstellar extinction. 
For more details on the catalog creation and validation see \citet{Drlica-Wagner:2021}.

A high-quality stellar sample is selected based on the \SExtractor quantity $\var{spread\_model}$ \citep{Desai:2012} measured in the DELVE \emph{g}-band. Specifically, we select objects with  $|\var{spread\_model\_g}| < 0.003$. 
The performance of this classifier was evaluated by matching sources in the DELVE DR1 catalog with the W04 HSC-SSP PDR2 catalog  \citep{hscpdr2:2019PASJ...71..114A}. 
For our analysis we choose a limiting magnitude of \CHECK{$g=23.1$ mag} where the stellar completeness drops to $\sim 60\%$ and contamination rapidly rises to $\sim 40\%$ as estimated from the HSC catalog.
For more information on morphological classification in this catalog see \citet{Drlica-Wagner:2021}.  
bright-end limit of 16th magnitude in the $g$-band is chosen to avoid saturation effects from bright stars. 
Additionally, since we are primarily interested in Main Sequence (MS) and Red Giant Branch (RGB) stars associated with old, metal-poor populations, we restrict the color range of our dataset to be $0.0 < (g - r)_0 < 1.0$. 
Only objects passing the above morphological, magnitude, and color cuts are used in the following matched-filter analysis. 

To account for missing survey coverage over our footprint, we quantify the sky area covered by DELVE DR1 in the form of \HEALPix maps. These maps account for missing survey coverage, gaps associated with saturated stars and other instrumental signatures. They are created using the  \code{healsparse}\footnote{\url{https://github.com/LSSTDESC/healsparse}} tool developed for the Legacy Survey of Space and Time (LSST) at the Vera C.\ Rubin Observatory, and its DECam implementation \code{decasu}\footnote{\url{https://github.com/erykoff/decasu}} to pixelize the geometry of each DECam CCD exposure. 
The coverage is calculated at a high resolution (\nside=16384; $\roughly 0.01 \amin^2$) and degraded to give a fraction of the lower resolution pixel area that is covered by the survey.

\subsection{\Gaia cross-match with DELVE DR1}
\label{sec:gaiaxmatch}
To enable a characterization of the proper motion of the Jet stream, we use the \Gaia EDR3  dataset \citep{GaiaeDR3:2020arXiv201201533G}. 
We begin by performing an angular cross match between the \Gaia EDR3 dataset and DELVE DR1 with a matching radius of 0\farcs5. 
This results in a catalog containing $\roughly 143$ million sources.   
Subsequently, a number of quality cuts are applied. 
Nearby sources are removed by applying a parallax cut similar to \citet{Pace:2019} of $\varpi - 3 \sigma_\varpi < 0.05$. 
To remove sources with bad astrometric solutions we place a cut on the renormalized unit weight error (\code{ruwe}) of $\code{ruwe} < 1.4$.
Then, a cut on BP and RP excess is applied following equation 6 of \citet{Riello:2020} ($|C^\star| < 3 \sigma_{C^\star})$. 
Additionally, only sources with $\code{astrom\_chi2\_al} < 2$ are kept to avoid sources with bad astrometric fits in \Gaia. 
We check that no known Active Galactic Nuclei (AGN) are in our sample by removing all sources that appear in the \Gaia table \code{gaiaedr3.agn\_cross\_id}. 
Finally, we remove faint sources with $G > 20$ mag to avoid contamination from stars with low signal-to-noise proper motion measurements.  
The resulting catalog is used for the analyses described in Sections \ref{sec:gradient} and \ref{sec:proper_motion}.

\section{Methods and Analysis}
In this section we describe our procedure to fit the track, distance gradient, proper motion, and morphology of the Jet stream.
We begin by performing an initial matched-filter selection for the Jet stream assuming the best-fit isochrone parameters and distance modulus from J18 (Section \ref{sec:initialmf}).
This allows us to determine an initial estimate for the Jet stream track. 
We then select candidate BHB stars that lie along the track and are clustered in proper motion space 
to determine a distance gradient as a function of angular distance along the stream (Section \ref{sec:gradient}).
Then we create a new optimized matched-filter map of the Jet stream using the distance gradient of the candidate BHB stars, and refitting an isochrone to a Hess difference diagram (Section \ref{sec:hess}).
This map is fit with a spline-based generative model to quantitatively characterize the track, intensity and width of the stream as a function of angular distance along the stream (Section \ref{sec:morphology}).
Finally, we select RGB and BHB stars consistent with being members of the Jet stream and fit a two component Gaussian mixture model to the selected stars determining the proper motion for the Jet stream including a linear gradient term (Section \ref{sec:proper_motion}). 
\label{sec:methods}

\begin{figure}[t!]
    \centering
    \includegraphics[width=0.98\columnwidth]{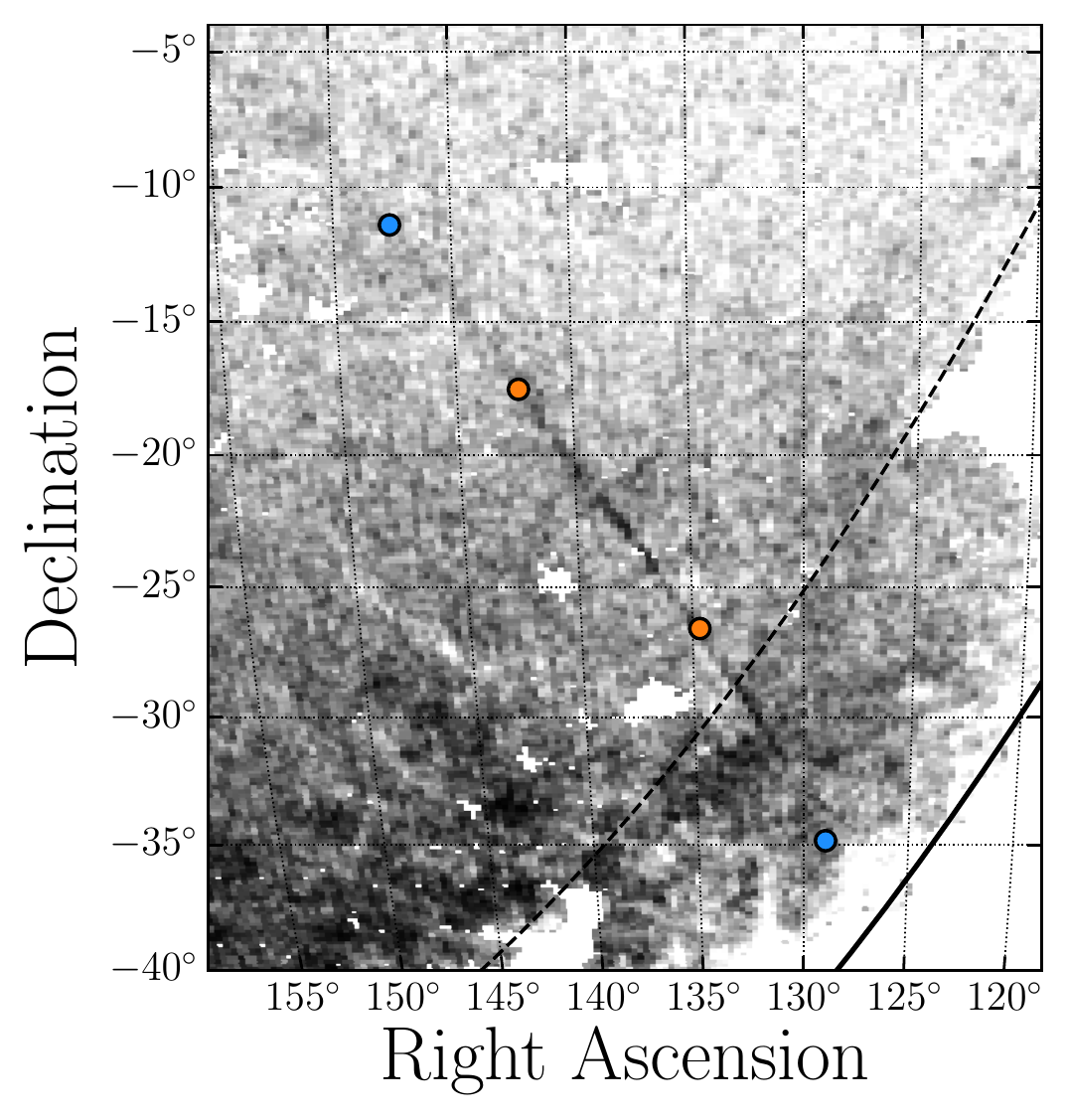}
    
    \caption{A stellar density map from the DELVE photometry showing the Jet stream at a distance modulus of $m-M=17.28$ ($\text{D}_\odot=28.6 \kpc$). The orange points show the extent of the stream identified by \citet{JethwaJetstream:2018}, and the blue points denote the new extent of the stream as detected in MSTO and BHB stars as determined in this study. The solid black line in the bottom right indicates the plane of the Milky Way ($b=0\degree$), and the dashed line shows $b=10\degree$.}
    \label{fig:initialmf}
\end{figure}
\subsection{Initial matched-filter Search}
\label{sec:initialmf}

To investigate the Jet stream in DELVE DR1, we began by applying a matched-filter algorithm in color--magnitude space similar to \citet{Shipp:2018,Shipp:2020}. 
The matched-filter is derived from a \citet{Dotter:2008} synthetic isochrone as implemented in \code{ugali} \citep{Drlica-Wagner:2020}.\footnote{\url{https://github.com/DarkEnergySurvey/ugali}}  Candidate MS stars are selected within a range of colors around the isochrone (Equation 4 in \citealt{Shipp:2018}) taking into account photometric uncertainty.
To select stars consistent with the Jet stream, we create a matched-filter based on the best-fit parameters (including distance modulus) taken from J18: an age of 12.1 Gyr, a metallicity of $\text{[Fe/H]}=$\fehjet, and a distance modulus of $m - M = 17.28 \magn$. 

Our selection is conducted using the DELVE DR1 catalog described in Section \ref{sec:data}. 
Stars are selected using the matched-filter and then objects are pixelized into \healpix pixels with \nside = 512 (pixel area of $\roughly 0.01 \deg^2$). 
The pixelized filtered map is corrected by the \response{geometric} survey coverage fraction for each pixel to account for survey incompleteness. \response{This is done by dividing the number of counts in each pixel by the observed fraction of that pixel \citep{Shipp:2018}. The coverage maps are created using the method described in Section 4.4 of \citet{DESCollaboration:2021},} and pixels with a coverage fraction less than 0.5 are removed from the analysis. 
 
\figref{initialmf} shows the results of this matched-filter selection. 
The Jet stream can be clearly seen to extend beyond the initial discovery bounds (marked by orange circles) in both directions.
At high declination the stream becomes fainter and more diffuse and appears to fan out, and at lower declination an additional prominent component can be seen with obvious density variations. 

\begin{figure*}[ht]
    \centering
    \includegraphics[width=0.98\textwidth]{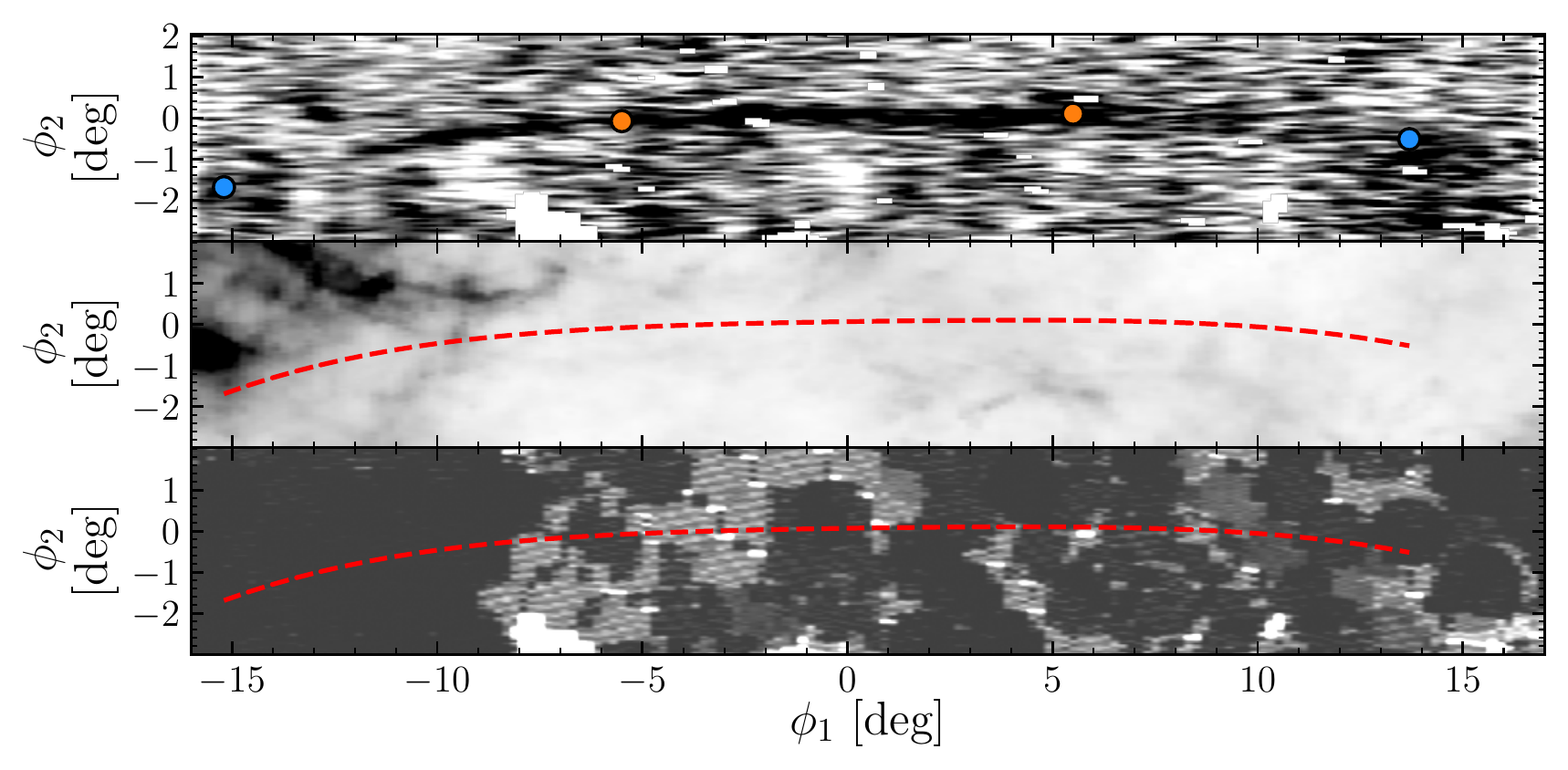}
    
    \caption{\emph{Top:} Jet stream density after applying the same matched-filter as Figure \ref{fig:initialmf}. Additionally, deviations from great circle path $\phi_2=0\degree$ are clearly seen. The orange points show the extent of the stream identified by \citet{JethwaJetstream:2018}, and the blue points denote the new extent of the stream as detected in MSTO and BHB stars. For the range $\phi_1 < -12.7\degree$ or $> 10 \degree$ the stream is only detected using BHB stars. 
    \emph{Middle:} SFD dust map of the same region.
    \response{\emph{Bottom:} Survey coverage map where the color shows the fraction of each pixel that is observed in the \delve survey. White corresponds to observed fractions lower that 0.5 which are excluded from our analysis, and the greyscale ranges from light to dark with observed fractions of 0.7 to 1.} The red dashed lines show the track of the Jet stream. }
    \label{fig:initialmf_jetframe}
\end{figure*}

In absence of an obvious stream progenitor, we choose the stream-centered coordinate frame to be the same as J18 defined by pole $(\alpha_{pole}, \delta_{pole}) =64.983\degree, 34.747\degree$ and a $\phi_1$ center of $\phi_1=63\degree$ ($\phi_1,\phi_2=0\degree,0\degree$ at $\alpha,\delta=138.62 \degree,22.10 \degree$). 
We define the rotation matrix to convert $\alpha, \delta$ to $\phi_1, \phi_2$ to be:
    \begin{equation}
        \boldsymbol{R}=
            \begin{bmatrix}
           -0.69798645 &  0.61127501 & -0.37303856 \\
            -0.62615889 & -0.26819784 &  0.73211677 \\
             0.34747655 &  0.74458900 &  0.56995374
        \end{bmatrix}.
    \end{equation}
The top panel of \figref{initialmf_jetframe} shows the transformed matched-filter stellar density map. 
This map has been smoothed by a Gaussian kernel with a size of {$\sigma = 0.06 \degree$}, and each column has been normalized to have the same median to correct for variable background stellar density along the field.
The track of the stream clearly deviates from the great circle path defined by $\phi_2=0\degree$. We fit a fourth-order polynomial to the peak intensity of the stream at each $\phi_1$ for the range $-14 \degree < \phi_1 < 14 \degree$ giving the following relation for $\phi_2$ as a function of $\phi_1$:
\begin{linenomath}
    \begin{align}
        \label{eqn:initial_track}
        \phi_2(\phi_1)=&\initaltrackA + \initaltrackB \times (\phi_1) \initaltrackC \times (\phi_1)^2 \\
        &+ \initaltrackD \times (\phi_1)^3 \initaltrackE \times (\phi_1)^4. \nonumber
    \end{align}
\end{linenomath}
The \response{middle} panel of Figure \ref{fig:initialmf_jetframe} shows the \citet[SFD]{Schlegel:1998} dust map in the transformed frame of Jet, \response{and the bottom panel shows the survey coverage map in this region.   
These two maps demonstrate that the detection of the Jet stream does not correlate with any linear extinction features or variations in survey coverage.} 

\subsection{Distance Gradient}
\label{sec:gradient}

Using the DELVE DR1 catalog cross-matched with \Gaia EDR3 (\secref{gaiaxmatch}), we identify candidate BHB stars and use them to measure a distance gradient along the Jet stream.
BHB stars are useful for determining a distance gradient because of the tight color-luminosity relation that allows for distance estimates with  $\roughly 10\%$ uncertainty to an individual BHB star \citep{Deason:2011}. 
To determine a distance gradient we use a similar method to \citet{Li:2020} who use BHB stars to measure the distance gradient of the ATLAS-Aliqa Uma stream. 
For the Jet stream, the gradient derived from BHB stars can be used to refine the matched-filter selection from \secref{initialmf}, make a reflex-corrected proper motion measurement in \secref{proper_motion}, and improve dynamical modeling of the stream (\secref{modeling}). 
Hereafter, all proper motions ($\mu_{\phi_1}^\star,\mu_{\phi_2}$) are assumed to be reflex corrected unless explicitly stated otherwise.
We select probable BHB stars along the track of the stream using a few criteria.
Initially we select all sources with $-12\degree < \phi_1 < 10 \degree$ and separation from the stream track $\Delta\phi_2 < 0.5 \degree$ ( Equation \ref{eqn:initial_track}). 
Then, a color cut is applied to select blue stars keeping only sources with $(g-r)_0$ between $-$0.3 and 0.0 mag. 
We then cut all sources with $g$-band magnitudes less than 17.0 mag or greater than 18.5 mag to reduce contamination from the Milky Way foreground.
For each candidate we derive an estimate of its distance modulus, $m-M$, by assuming it is a BHB star and using the relation for $M_g$ vs.\ $(g-r)_0$ from \citet{Belokurov:2016}.

\begin{figure*}[ht]
    \centering
    \includegraphics[width=\textwidth]{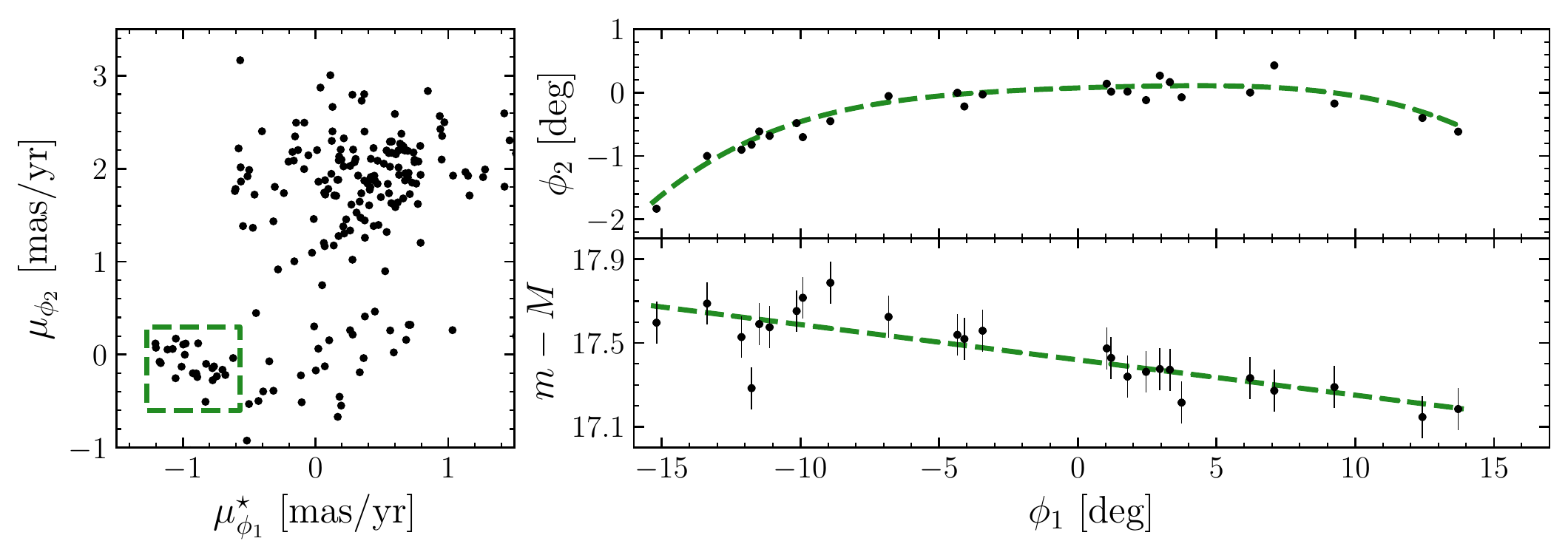}
    \caption{
    \emph{Left:} proper motions of candidate BHB stars along the Jet stream. The green box shows our proper motion selection for likely members. \response{Contributions from the Milky Way foreground and stellar halo are seen at $(\mu_{\phi_1}^\star,\mu_{\phi_2})\sim (0.5,2)$ and $(0.5,0)$ mas/yr respectively}
    \emph{Right:} on-sky distribution (top) and distances modulus (bottom) of candidate BHB stars that are likely associated with the Jet stream. The matched-filter stream track (Equation \ref{eqn:initial_track} is shown as green dashed line on the top plot, and the linear fit on distance modulus,  $(m − M)$, of the candidate BHB stars (Equation \ref{eqn:grad}) is shown as a green dashed line on the bottom plot.}
    \label{fig:bhb_grad}
    \label{fig:pm_bhb}
\end{figure*}

To further remove contaminant stars from the Jet stream BHB stellar sample, we use the \Gaia EDR3 proper motions of candidate BHB stars along the Jet stream track.
We use the distance estimated for each BHB star to correct their proper motions for the solar reflex motion assuming a relative velocity of the Sun to the Galactic standard of rest to be $(U_\odot,V_\odot,W_\odot)=(11.1,240.0,7.3) \text{ km}\text{/s}$ \citep{bovy2012:ApJ...759..131B}.
\figref{pm_bhb} (left panel) shows the resulting measured proper motion of our BHB candidate sample. 
The proper motion signal of the Jet stream is seen at $(\mu_{\phi_1}^\star,\mu_{\phi_2})\roughly(-1,0)$ mas/yr, where we define $\mu^\star_{\phi_1}=\mu_{\phi_1}\cos(\phi_2)$. \response{Additionally, contributions due to the Milky Way foreground and stellar halo are seen at $(\mu_{\phi_1}^\star,\mu_{\phi_2})\roughly(0.5,2)$ and $(0.5,0)$ mas/yr respectively} 
The BHB candidates within the green box are selected as likely members of the Jet stream to be used to estimate the distance gradient. 
\figref{bhb_grad} (right panels) shows on-sky positions and distances of the likely member candidate BHB stars. 

Using these derived distances and assuming an uncertainty of 0.1 mag on the distance modulus for each BHB star \citep[]{Deason:2011}, we fit for the distance modulus as a function of position along the stream using a simple linear fit. 
We find the following relation for the distance modulus as a function of $\phi_1$ along the Jet stream,
\begin{equation}
    \label{eqn:grad}
    (m-M)=\dmtrackA \dmtrackB \times (\phi_1).
\end{equation}
The heliocentric distance of the Jet stream is found to vary from \CHECK{$26 \kpc$} to \CHECK{$34.5 \kpc$} over its observed length with a gradient of $-0.2$ kpc/deg.

\subsection{Creation of optimized matched-filter map}
\label{sec:hess}
\begin{figure}
    \centering
    \includegraphics[width=\columnwidth]{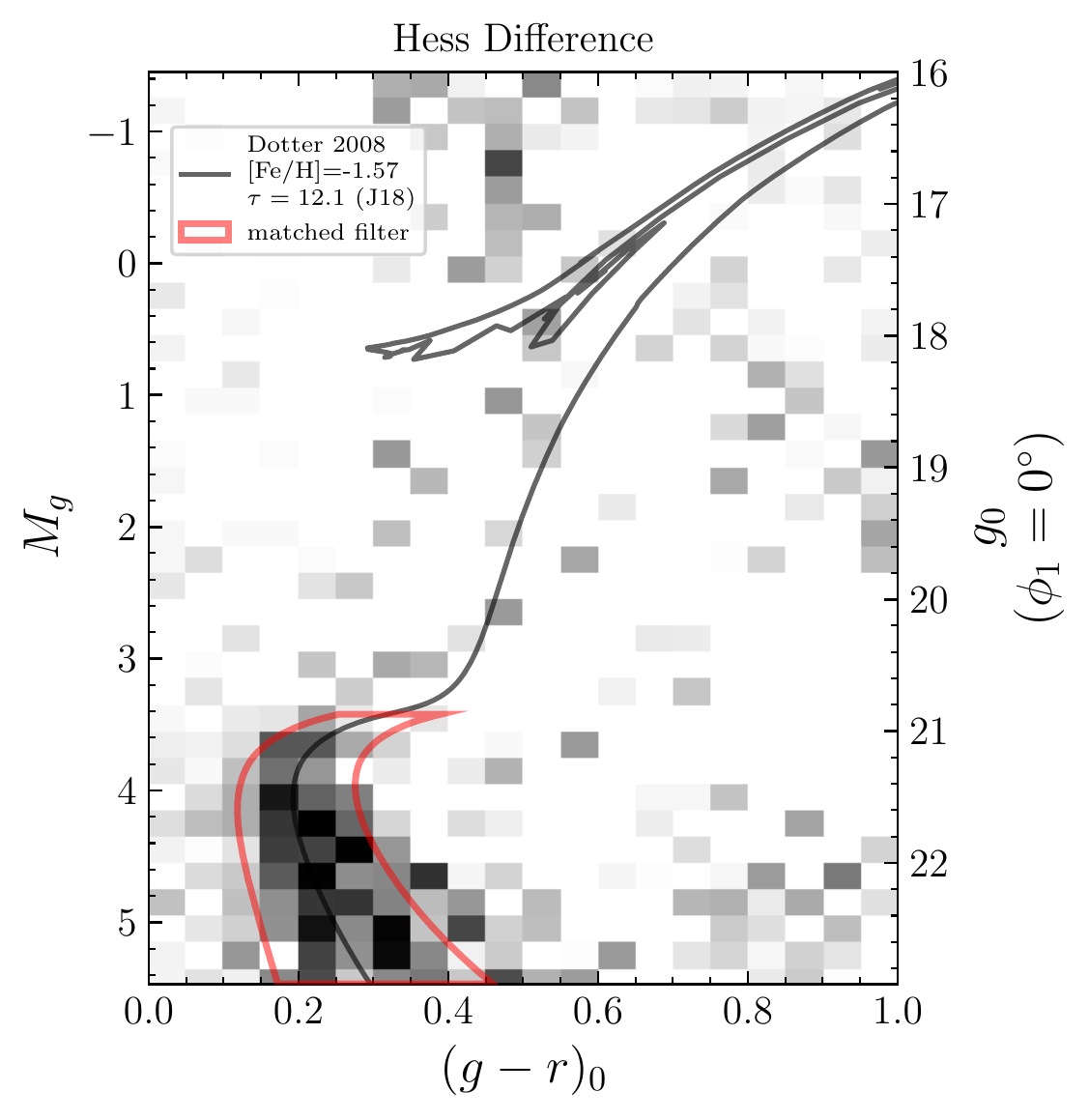}
    \caption{Hess difference diagram created by subtracting a background region from the on-stream region. The main sequence of the Jet stream is clearly seen. The \citet{Dotter:2008} isochrone we use for our analysis with a metallicity of [Fe/H]=\fehjet\xspace and age of $\tau=12.1 \Gyr$ is shown as a solid black line. The red line shows the matched-filter used to create the optimized map shown in the top row of Figure \ref{fig:2dmodel}. The left y-axis shows the absolute magnitude of sources and the right y-axis shows the corresponding apparent magnitude at $\phi_1=0\degree$.}
    \label{fig:hess}
\end{figure}

We next use our measured distance gradient to find a best-fit isochrone, and create an optimized matched-filter map to study the morphology of the stream. 
We begin by creating a Hess difference diagram shown in Figure \ref{fig:hess}.
This is done by selecting stars along the stream track with $-11 \degree < \phi_1 < -8\degree$ as well as stars with $-5 \degree < \phi_1 < 6 \degree$, excluding the area around the observed under-density in the stream at $\phi_1 \sim -6.5\degree$ to increase signal-to-noise.
In $\phi_2$ we select stars within 2 times the observed width of the stream ($w$) as a function of $\phi_1$ (the width is derived in Section \ref{sec:morphology}). 
At $\phi_1=0$ we find a width of $w=0.16\degree$ and this value varies from $w=0.13\degree$ at $\phi_1=-11\degree$ to $w=0.19\degree$ at $\phi_1=6\degree$.
Additionally, a background region is selected to be along the same $\phi_1$ range but above and below the stream, with  $1\degree < |\phi_2| < 2\degree$. 
For the stars in each of these regions we compute the absolute $g$-band magnitude ($M_g$) assuming a distance modulus derived from the observed BHB stars' gradient (Equation \ref{eqn:grad}). 
Then we select only stars with $M_g < 5.47$; this corresponds to the faint limit of our catalog ($g_0=23.1$) at $\phi_1=-13\degree$, the most distant portion of the detected stream. 
This absolute magnitude selection ensures that the observed density variations in the matched-filter map are not affected by the survey completeness. 
We then create binned color--magnitude diagrams (CMDs) for the on-stream and background selections and subtract the background from the on-stream region correcting for relative areas. 
The result of this process is shown in Figure \ref{fig:hess}.

Next we fit an isochrone to this Hess difference diagram using a similar methodology to \citet{Shipp:2018}.
Briefly, we model the observed binned CMD of the stream region as a linear combination of the background region and a stellar population following a \citet{Dotter:2008} isochrone. 
Then we compute the likelihood that the binned CMD is a Poisson sample of the model. 
This likelihood is then sampled using \code{emcee} \citep{Foreman_Mackey:2013}.
We find a best-fit isochrone consistent with the J18 result and distance modulus in strong agreement with the distance derived from candidate BHB stars  ($(m-M)_{MS}-(m-M)_{BHB}=0.01^{+0.05}_{-0.05}$).

Since we find consistent results, we use the J18 isochrone to create an optimized matched-filter that is used in Section \ref{sec:morphology} to quantitatively characterize distance variation. 
This map covers the region defined by $-16 \degree\leq \phi_1 \leq 17 \degree$, and $-3 \degree\leq \phi_1 \leq 1 \degree$. We choose a pixel size of 0.2 deg in $\phi_1$ and 0.05 deg in $\phi_2$. 
The distance modulus of the matched-filter follows Equation \ref{eqn:grad}. 
The result of this more optimal matched-filter is shown in the top panel of \figref{2dmodel}. 
We note that while the image in \figref{2dmodel} has been smoothed with a 0.08 deg Gaussian kernel, we do not apply any smoothing when fitting our model to the stream data.

\subsection{Fitting Stream Morphology}
\label{sec:morphology}
To quantitatively characterize the observed features of the stream morphology we use a generative stream model developed by \citet{koposov2019:orphanMNRAS.485.4726K} and \citet{Li:2020}, which is  similar to that of \citet{Erkal2017:pal5MNRAS.470...60E}.
This model uses natural cubic splines with different numbers of nodes to describe stream properties. 
This model is implemented in \code{STAN} \citep{Carpenter:2017} and is fit to the data using the Hamiltonian Monte Carlo No-U-Turn Sampler (NUTS) to efficiently sample the high dimensional parameter space.

The stream is modeled by a single Gaussian in $\phi_2$ with central intensity, width, and track that are allowed to vary as functions of $\phi_1$.
The parameters of the model are $\mathcal{I}(\phi_1)$, $w(\phi_1), \, \Phi_2(\phi_1), \, \mathcal{B}_1(\phi_1), \, \mathcal{B}_2(\phi_1),$ and $\mathcal{B}_3(\phi_1)$, which describe the logarithm of the stream central intensity, the logarithm of the stream width, the stream track, the log-background density, the slope of the log-background density, and the quadratic term of the log-background density, respectively. \response{For more details see \citet[]{koposov2019:orphanMNRAS.485.4726K} (Section 3.1 and equation 2)}
The model is fit to the binned matched-filter data described above using Equation \ref{eqn:grad} to describe the distance modulus as a function of $\phi_1$. 
We assume the number of stars in an individual pixel of the  matched-filter map is a Poisson sample of the model density at that location. 

Following \citet[]{Li:2020}, we use Bayesian optimization to determine model complexity in a data-driven way. In particular, the number of nodes for all parameters except the stream width \response{are hyper-parameters of this model, and they are determined through model selection. Where Bayesian optimization \citep{Gonzalez2016,gpyopt:2016} of the cross-validated ($k=3$) log-likelihood function is used.} 
For the parameters [$\mathcal{I}(\phi_1)$, $w(\phi_1)$, $\Phi_2(\phi_1)$,  $\mathcal{B}_0(\phi_1)$, $\mathcal{B}_1(\phi_1)$, $\mathcal{B}_2(\phi_1)$] we find the optimal number of nodes to be [11, 3, 8, 28, 25, 5], respectively. 
For each parameter, the range of allowable nodes is 3 to 30 except for the width, $w(\phi_1)$, and quadratic term of the log-background density, $\mathcal{B}_3(\phi_1)$, which have their maximum number of nodes constrained to 15 and 10, respectively.
The model is run for 1500 iterations with the first 700 discarded as burn in. 

\begin{figure}[ht]
    \centering
    \includegraphics[width=0.98\columnwidth]{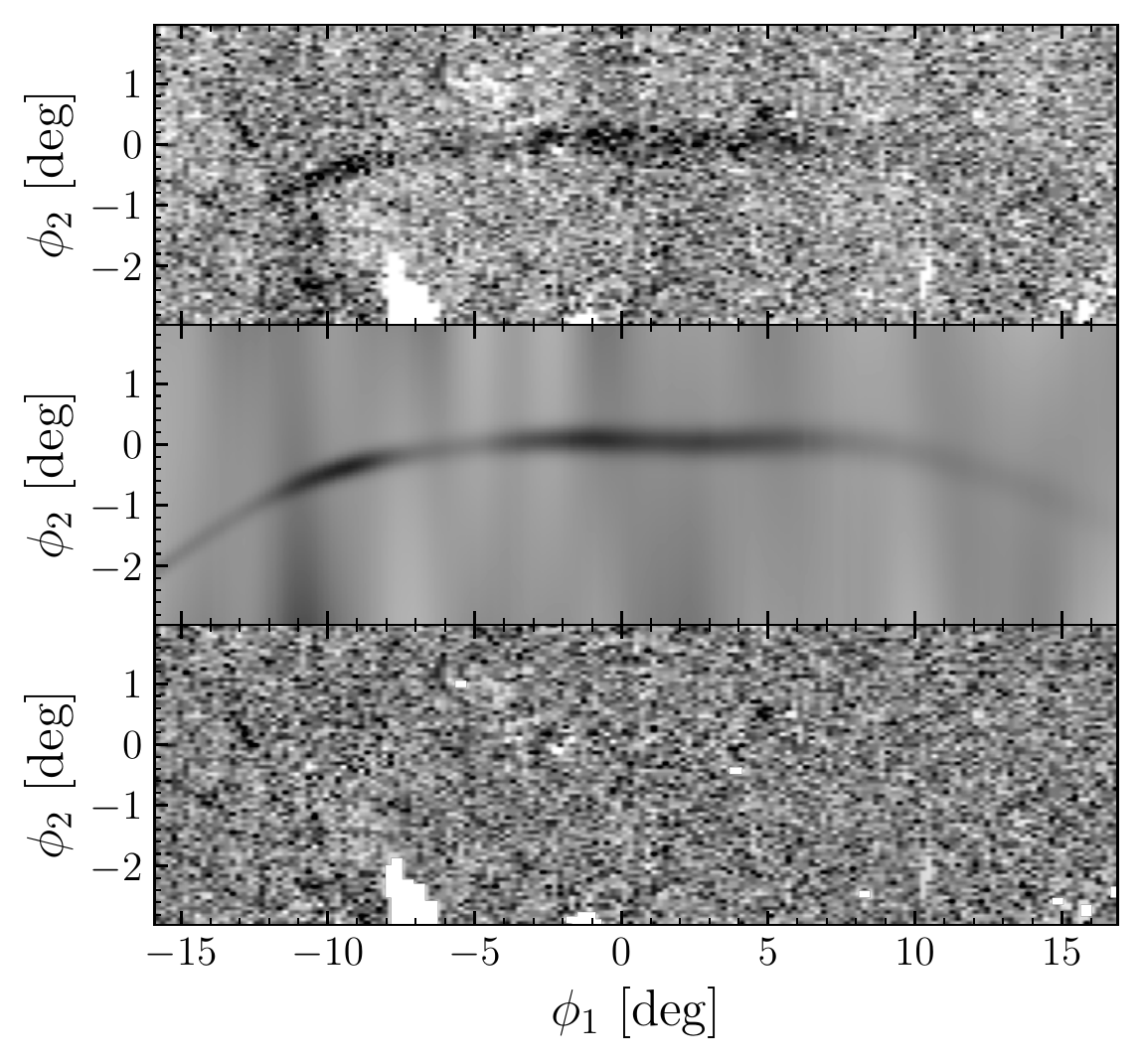}
    \caption{Modeling the track, width, and intensity of the Jet stream from \delve DR1 photometric data. \emph{Top:} The density of stars that pass the optimized matched-filter selection that takes into account the observed distance gradient in the region of the Jet stream.  \emph{Middle:} The maximum {\it a posteriori} (MAP) model of the data shown in the top panel containing both stream and background components.  \emph{Bottom:} The residual density map showing the observed density minus the model.}
    \label{fig:2dmodel}
\end{figure}

\begin{figure}[ht]
    \includegraphics[width=0.97\columnwidth]{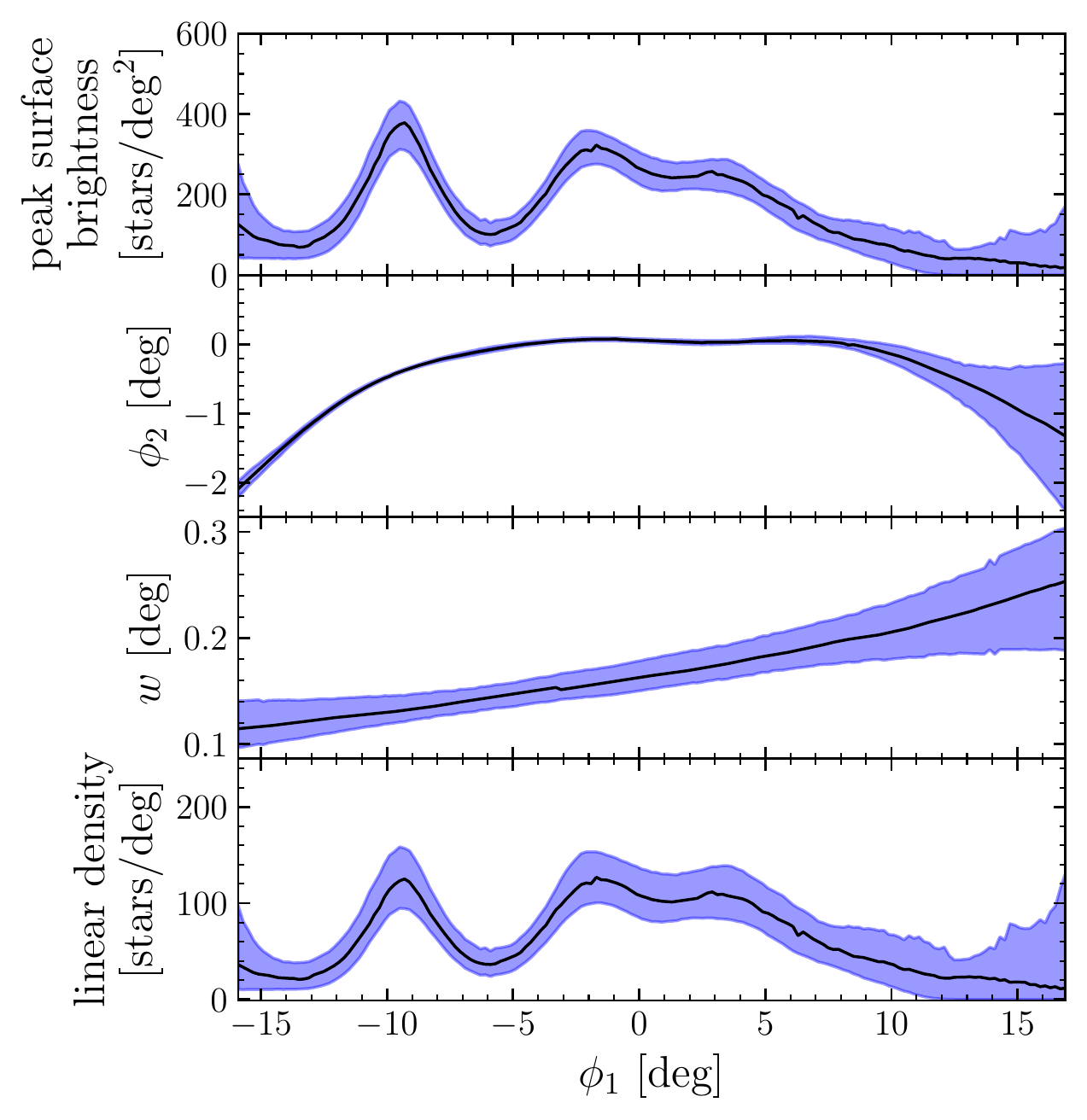}
    \caption{Measurement of Jet stream parameters as a function of position along the stream ($\phi_1$) as derived from modeling the DELVE DR1 stellar density maps.
    From top to bottom are stream surface brightness, 
    stream track, stream width, and linear density. The shaded area shows the 68\% containment peak interval, and the black line shows the best-fit estimate of each parameter. 
    }
    \label{fig:1dmodel}
\end{figure}

The results of the model fit are shown in Figures \ref{fig:2dmodel} and \ref{fig:1dmodel}.
For Figure \ref{fig:2dmodel} the top row shows the observed matched-filter map, the best-fit model is shown on the second in the same color scale, and the residual of the model subtracted from the data is shown on the bottom row. 
The key features captured by this model are the variations in the density of stars. 
A large gap can be seen at $\phi_1=-6 \degree$, and peaks in the intensity are found at at $\phi_1=-9 \degree, -2.5 \degree$ and $4 \degree$. 
The model does not capture all of the observed small-scale substructure. 
In particular, the off-track structure seen crossing the stream at $\phi_1=-12$ or the overdensity above the stream at $\phi_1=5 \, \degree$ are discussed in more detail in Section \ref{sec:discsmall}.

The nature of the on-stream structure can be better evaluated by looking at the extracted stream parameters in Figure \ref{fig:1dmodel}. These plots show the stream surface brightness, the on-sky track, stream width and linear density. In each panel, the best-fit value calculated as the maximum \emph{a posteriori} (MAP) of the posterior for each parameter as a function of $\phi_1$ is shown as a black line, and the 68\% containment peak interval is shown as the blue shaded region.
The apparent width of the stream increases with $\phi_1$, consistent with expected projection effects due to a constant width and the observed distance gradient.
This is supported by the relatively constant linear density over large scales.

\subsection{Proper Motion of the Jet Stream}
\label{sec:proper_motion}
In this section, we use the cross-matched DELVE DR1 and \Gaia EDR3 catalog to measure the proper motion of the Jet stream. 
In Section \ref{sec:gradient} we demonstrated that the proper motion signature of BHB stars could be clearly separated from the Milky Way foreground.
We seek to extend our analysis to the full stellar population of Jet, applying several additional physically motivated cuts to reduce Milky Way foreground contamination.
Then we perform a Gaussian mixture model fit with stream and Milky Way components to measure the proper motion of Jet. 

\subsubsection{Data Preparation}
\begin{figure*}
    \centering
    \includegraphics[width=0.98\textwidth]{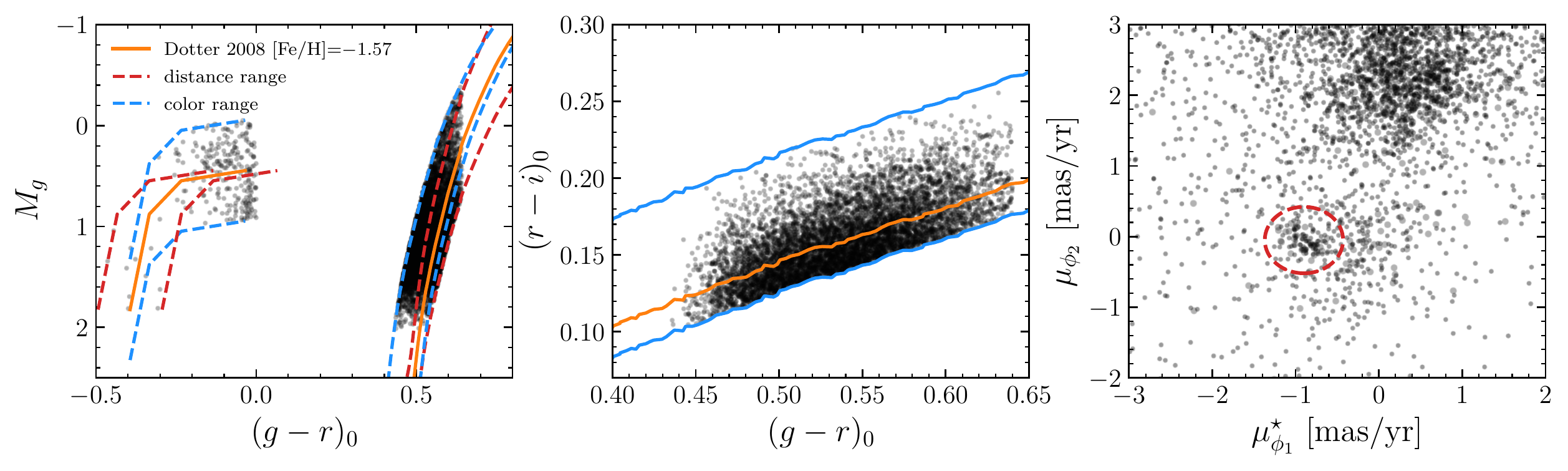}
    \caption{Selection of stars used to measure the proper motion of the Jet stream. \emph{Left:} A color--magnitude diagram demonstrating the selection applied to the data. The orange lines show the empirical M92 horizontal branch and Dotter isochrone ($[\text{Fe/H}]=$\fehjet) used to select data. The blue dashed lines show our selection range in distance, and the red dashed lines our selection range in color. 
    \emph{Center:}  A color-color plot showing our selection range in blue around an empirical stellar locus of dereddened DES photometry in orange. 
    \emph{Right:} Proper motion in stream coordinates of the sample of stars passing all selection cuts. The signal from the Jet stream is easily seen at $(\mu_{\phi_1}^\star,\mu_{\phi_2}) \sim (-1,0)$ mas/yr (circled in red).}
    \label{fig:pmcuts}
\end{figure*}

Starting with the stellar catalog from Section \ref{sec:gaiaxmatch}, we apply several cuts to reduce Milky Way contamination in our sample and to highlight the Jet stream population. These cuts are depicted visually in the left two panels of Figure \ref{fig:pmcuts}. 
The following selection process largely follows the methodology set out by \citet{shipp:2019ApJ...885....3S} and \citet{Li:2019}.

We begin by calculating the absolute magnitude in the $g$-band ($M_g$) for each star assuming a distance given by the fit to the BHB stars (Equation \ref{eqn:grad}; $m-M(\phi_1)$). 
A magnitude cut is made keeping only sources with $M_g < 2$ to remove faint sources with large proper motion uncertainties.   
Then, a color--magnitude selection is applied selecting stars in $(g-r)_0$ vs.\ $M_g$ color--magnitude space (Fig. \ref{fig:pmcuts} left panel). 
RGB stars are selected based on the \citet{Dotter:2008} isochrone used in Section \ref{sec:initialmf}. 
We select stars that meet either of the following conditions:
\begin{linenomath}
    \begin{align}
       -0.08 &\leq &(g-r)_0-&(g-r)_{\rm Dotter} &\leq 0.02\\ \nonumber
        -0.5 &\leq & g_0-&(g_0)_{\rm Dotter} &\leq 0.5. \nonumber
    \end{align}
\end{linenomath}
where $(g-r)_{\rm Dotter}$ is the isochrone color at a given observed magnitude and $(g_0)_{\rm Dotter}$ is the isochrone magnitude at a given observed color. 
Next, we applied a $(g-r)_0$ vs.\ $(r-i)_0$ color--color cut to select metal poor stars based on an empirical stellar locus that is derived from dereddened DES data \citep[second panel;][]{Pace:2019,Li:2018}. 
This locus gives the median $(r-i)_0$ colors for each $(g-r)_0$ bin, For each star we compute $\Delta_{ri}=(r-i)_{0}-(r-i)_{\rm med}$ where $(r-i)_{0}$ is the observed color of a star and $(r-i)_{\rm med}$ is the median $(r-i)_0$ color of stars with the same $(g-r)_0$ color taken from the empirical stellar locus. 
Only stars within $-0.02 < \Delta_{ri} < 0.1$ are kept.  
Finally, a spatial cut is applied only keeping stars within $3 \times w(\phi_1)$ of the stream track where $w(\phi_1)$ is the stream width and track taken from the modeling in Section \ref{sec:morphology}.

To select candidate horizontal branch members for the proper motion fit, we use an empirical horizontal branch of M92 initially derived in  \citet{Belokurov:2007} and transformed to the DES photometric system \citep{Li:2019,Pace:2019}. 
We select stars with $(g-r)_0$ colors within $\pm0.1$ mag and $M_g$ within $\pm0.5$ mag of the empirical horizontal branch.   

After applying these cuts, we perform a reflex correction on the proper motion measurements of the remaining sources (assuming the distance fit from Equation \ref{eqn:grad}). 
The proper motion signal of Jet is easily identified in the third panel of Figure \ref{fig:pmcuts} as the overdensity at $(\mu_{\phi_1}^\star,\mu_{\phi_2})\sim (-1,0)$\, mas/yr. 
To quantitatively measure the proper motion and any gradient in the proper motion, we use a Gaussian mixture model based analysis described in the following section.

\subsubsection{Mixture Model}
\label{sec:mixture}
To determine the proper motion of the Jet stream, we use a simple mixture model consisting of Gaussian distributions for the stream and Milky Way foreground. 
The fit is performed on the candidate RGB and BHB stellar sample defined in the previous section and shown in the right panel of Figure \ref{fig:pmcuts}.
The likelihood and fitting methodology follows that of \citet{Pace:2019} and \citet{shipp:2019ApJ...885....3S}.
The complete likelihood is given by:
\begin{equation}
    \mathcal{L}= (1-\lambda)\mathcal{L}_{\rm Jet}+\lambda \mathcal{L}_{\rm MW}
\end{equation}
where $\mathcal{L}_{\rm Jet}$ is the likelihood that a star belongs to the Jet stream component, and $\mathcal{L}_{\rm MW}$ refers to the likelihood that a star belongs to the Milky Way foreground component.
The fraction of stars that belong to the Milky Way component is denoted by $\lambda$.
Each likelihood term is made up of the product of both spatial and proper motion likelihoods, 
\begin{equation}
    \mathcal{L}_{\rm Jet/MW}=\mathcal{L}_{\rm spatial}\; \mathcal{L}_{\rm PM}.
\end{equation}
For the stream spatial component we use the results from Section \ref{sec:morphology} for the stream track, $\Phi_2(\phi_1)$, and width, $w(\phi_1)$, assuming the stream follows a Gaussian distribution in $\phi_2$ around this track (i.e. $\phi_{2,\text{obs}}-\Phi_2(\phi_1)$ with standard deviation $w(\phi_1)$. 
The Milky Way spatial component is assumed to be uniform. 
\begin{equation}
  \mathcal{L}_{\rm spatial} =
  \begin{cases}
       \mathcal{N}(\phi_2\given\Phi_{2}(\phi_1), w_\Jet)& \text{Jet component} \\
       \mathrm{Uniform} & \text{MW component} \\
  \end{cases}        
\end{equation}
The proper motion term of the likelihood is modeled by the combination of several multivariate Gaussians. 

Each Gaussian is defined to have a mean proper motion vector $\boldsymbol{\chi}$ given by $\boldsymbol{\chi}=(\mu_{\phi_1, Jet}^\star(\phi_1),\mu_{\phi_2, Jet}(\phi_1))$. We model both components of $\boldsymbol{\chi}$ as a linear functions of $\phi_1$. The covariance, $\boldsymbol{C}$, is defined by an observational error component and an intrinsic component:
\begin{linenomath}
\begin{align}
    \boldsymbol{C}&=\boldsymbol{C}_{obs}+\boldsymbol{C}_{intrinsic}\nonumber\\
    &=
    \begin{bmatrix}
    \epsilon_{\mu_{\phi_1}\, \cos \phi_2}^2 &
    \epsilon_{\mu_{\phi_1}\, \cos \phi_2 \times \mu_{\phi_2}}\\
    \epsilon_{\mu_{\phi_1}\, \cos \phi_2 \times \mu_{\phi_2}}    & 
    \epsilon_{\mu_{\phi_2}}^2 \
    \end{bmatrix} 
    + \\
    &\quad\;\begin{bmatrix}
    \sigma_{\mu_{\phi_1}\, \cos \phi_2}^2  &
     \sigma_{\mu_{\phi_1}\, \cos \phi_2} \times  \sigma_{\mu_{\phi_2}} \times \rho\\
     \sigma_{\mu_{\phi_1}\, \cos \phi_2} \times  \sigma_{\mu_{\phi_2}} \times \rho   & 
    \sigma_{\mu_{\phi_2}}^2\\
    \end{bmatrix}\nonumber
\end{align}
\end{linenomath}
where $\epsilon$ represents the proper motion errors, $\sigma$ is the intrinsic proper motion dispersions, and $\rho$ is a correlation term. 
For the stream component, the intrinsic dispersion is assumed to be $5 \km/\second$ which at the distance of the stream corresponds to $\sim 0.04 \mas/\yr$ and the correlation terms are assumed to be zero. 

The model then has 
9 free parameters: 
The systemic proper motions of the stream measured at
$\phi_1=0 \degree$ ($\Bar{\mu}_{\phi_1}\cos(\phi_2),\Bar{\mu}_{\phi_2}$), 
the proper motion gradients in each coordinate direction for the stream
($d\mu_{\phi_1}/d\phi_1,d\mu_{\phi_2}/d\phi_1$) in units of $\mas/ 10 \degree$, the mean proper motion of the Milky Way foreground Gaussian ($\mu_{\phi_1\, {\rm MW}},\mu_{\phi_2\,{\rm MW}}$), the dispersion of the Milky Way foreground Gaussian ($\sigma_{\phi_1\, {\rm MW}},\sigma_{\phi_2\,{\rm MW}}$), and the fraction of stars that belong to the stream component ($\lambda$).
The proper motion of the stream component as a function of $\phi_1$ is given by
\begin{linenomath}
\begin{align}
    \mu_{\phi_1, \Jet}(\phi_1)&=\Bar{\mu}_{\phi_1} +\, \frac{d\mu_{\phi_1}}{d\phi_1} \times (\phi_1/10 \deg)\\
    \mu_{\phi_2, \Jet}(\phi_1)&=\Bar{\mu}_{\phi_2} +\, \frac{d\mu_{\phi_2}}{d\phi_1} \times (\phi_1/ 10\deg) \nonumber.
\end{align}
\end{linenomath}

\noindent The total proper motion likelihood is then given by
\begin{equation}
    \mathcal{L}_{\rm PM}=
    \sum_{N=1}^k
    \norm
    ((\boldsymbol{\mu}_{\phi_1, obs}^{\star}, \boldsymbol{\mu}_{\phi_2, obs})_N
    \given\boldsymbol{\chi}_{true, N}\;,\boldsymbol{C_N}).
\end{equation}

Parameter inference is conducted using a Hamiltonian Monte Carlo No-U-Turn Sampler (NUTS) implemented in \code{STAN} \citep{Carpenter:2017}. We use 10 parallel chains with 2000 iterations each (1000 of the iterations are discarded as a burn in).
Convergence is verified using the Gelman-Rubin $\hat{R} < 1.1$ diagnostic \citep[]{Gelman:1992}.

The results of our fit are listed in Table \ref{tab:propermotion}. 
We find the results from the mixture model ~($\Bar{\mu}_{\phi_1,\Jet}^\star, \Bar{\mu}_{\phi_2,\Jet}) = (\pmMuAStr\pm \pmMuAStrUP,\pmMuBStr\pm \pmMuBStrUP)$~mas/yr, which agrees with our rough estimate from the observed overdensity of candidate BHB stars in \secref{gradient}. 
The $\Bar{\mu}_{\phi_2}$ value is near zero as expected for a stable stream that has not been heavily perturbed. 
We detect gradients in both proper motion coordinates that are similar in magnitude.
Based on these results, the tangential velocity of the Jet stream at $\phi_1=0\degree$ is $v_{tan}=195\pm3$ km/s.

A membership probability is calculated for each star by taking the ratio of the stream likelihood to the total likelihood: $p_{mem}=\lambda\mathcal{L}_\Jet/(\lambda\mathcal{L}_{\rm Jet}+(1-\lambda)\mathcal{L}_{\rm MW})$. 
To determine the value of $p_{mem,i}$ for each star we calculate $p_{mem}$ for each point in the posterior of our fit and take $p_{mem,i}$ to be the median of the calculated values for each star. 
A star is then considered a high (medium) probability member if $p_{mem,i} > 0.8\; (0.5)$. 
For our sample, $\highnumRGBcandpm\; (\mednumRGBcandpm)$ candidate RGB stars and $\highnumBHBcandpm\; (\mednumBHBcandpm)$ candidate BHB stars pass this criterion. 

\begin{figure*}
    \centering
    \includegraphics[width=0.98\textwidth]{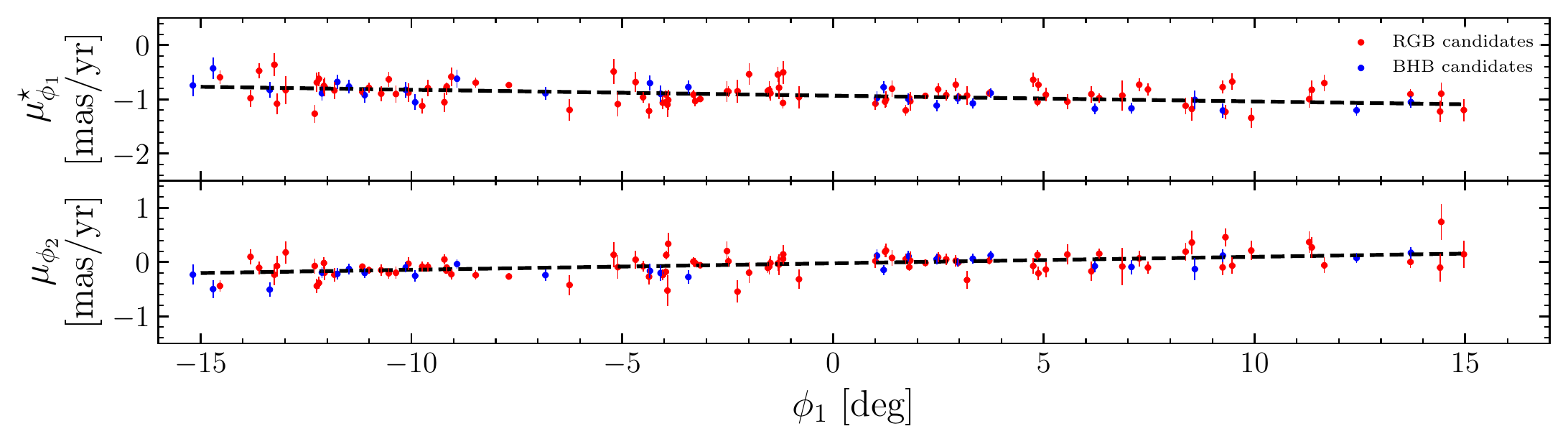}
    \caption{Proper motion of high probability member RGB (red) and BHB (blue) stars ($p_{mem} > 0.8$). Additionally the line of best-fit proper motion is shown as a black dashed line. All the BHB stars used to measure the distance gradient are also high probability members.}
    \label{fig:pm_result}
\end{figure*}

The proper motion of the high probability stars are shown in \figref{pm_result} along with lines showing the best-fit stream proper motion ($\Bar{\mu}_{\phi_1}^\star(\phi_1),\Bar{\mu}_{\phi_2}(\phi_1))$ from our analysis. 
The BHB stellar sample (blue points) is very similar to the sample selected by a rough cut in \figref{pm_bhb}, and it can be seen that these stars closely follow the proper motion gradient ($d\mu/d\phi_1$) found in our analysis. 
In Appendix \ref{sec:memprob}, we include Table \ref{tab:memprob} that contains the properties of all candidate BHB and RGB stars with membership probability ($p_{mem}$) higher than 10\%.  
With the current dataset we are only able to fit for linear evolution of the proper motion with $\phi_1$, but with future spectroscopic datasets can test for a more complex evolution of the proper motion as a function of $\phi_1$ (e.g., quadratic).

Previous studies such as \citet{shipp:2019ApJ...885....3S} have looked for signs of large scale perturbation of stellar streams from the influence of the LMC or Milky Way bar \citep{Li:2018, Erkal:2018, koposov2019:orphanMNRAS.485.4726K, Vasiliev:2021}. 
Evidence of these interactions sometimes appears as a mismatch between the proper motion of the stream ($\mu_{\phi_2}/\mu_{\phi_1}$) and the derivative of the stream track ($d\phi_2/d\phi_1$)~\citep{Erkal:2019}. 
In the case of Jet, we find that the ratio of the proper motions to the stream track ($\mu_{\phi_2}/\mu_{\phi_1})/(d\phi_2/d\phi_1$) has an average value of $2.00 \pm 1.18$ over the extent of the stream. 
This is consistent with a value of one which indicates the proper motions are largely aligned with the track of the stream. 

\begin{deluxetable}{l ccccc}
\tablecolumns{13}
\tablewidth{0pt}
\tabletypesize{\small}
\tablecaption{ Results of the proper motion fit in reflex corrected proper motion coordinates}
\label{tab:propermotion}
\tablehead{
\colhead{Parameter} & \colhead{Value} & \colhead{Prior} & \colhead{Range}   & \colhead{Units}}
\startdata
$\Bar{\mu}_{\phi_1}^\star$      & $\pmMuAStr\pm \pmMuAStrUP$    & Uniform & -10,10     & mas/yr\\
$\Bar{\mu}_{\phi_2}$            & $\pmMuBStr\pm \pmMuBStrUP$    & Uniform & -10,10     & mas/yr\\
$d\mu_{\phi_1}^\star/d\phi_1$   & $\dpmMuAStr\pm \dpmMuAStrUP$  & Uniform & -3,3 & mas/yr/10 deg\\
$d\mu_{\phi_2}/d\phi_1$         & $\dpmMuBStr\pm \dpmMuBStrUP$  & Uniform & 3,3  & mas/yr/10 deg\\
$\lambda$                       &$\fStr \pm \fStrUP$             & Uniform & 0,1 &\\
\enddata
\end{deluxetable}

\section{Dynamical Modeling}
\label{sec:modeling}
Using our measurements of the stream track, distance, and proper motion, we can fit a dynamical model to the data. 
The Jet stream is modeled in the same method as \citet{Erkal:2019} and \citet{Shipp:2021}.
We make use of the modified Lagrange Cloud Stripping (mLCS) technique developed in \citet{Gibbons:2014} adapted to include the total gravitational potential of the stream progenitor, the Milky Way, and the LMC. 
Following \citet{Erkal:2019}, the Milky Way and LMC are modeled as independent particles with their respective gravitational potentials which allows us to capture the response of the Milky Way to the LMC. 
The Milky Way potential is modeled with 6 axisymmetric components, namely bulge, dark matter halo, thin and thick stellar disk, and HI and molecular gas disk components following \citet{McMillan:2017}. 
Following \citet{Shipp:2021}, we normalize the Milky Way potential to the realization of the \citet{McMillan:2017} potential that yields the best-fit from the ATLAS data  \citep[$M_{\rm MW} = 8.3 \times 10^{11} M_{\rm \odot}$;][]{Li:2020}.
We evaluate the acceleration from the potential using \code{galpot} \citep{Dehnen:1998}. 
We take the Sun's position ($R_0=8.23$ kpc) and 3D velocity, ($\text{U}_\odot,\text{V}_\odot,\text{W}_\odot)=(8.6,232.8,7.1)$ km/s, from \citet{McMillan:2017}. 

We model the mass distribution of the LMC as a stellar disk and a dark matter halo. The stellar disk is modelled as a Miyamoto-Nagai disk \citep{Miyamoto:1975} with a mass of $3\times10^9 M_\odot$, a scale radius of $1.5$ kpc, and a scale height of $0.3$ kpc. The orientation of the LMC disk matches the measurement of \cite{vanderMarel:2014}. The LMC's dark matter halo is modelled as a Hernquist profile \citep{Hernquist:1990}. 
We fix the total infall mass of the LMC to $1.5 \times 10^{11} \Msun$, consistent with the value derived in \citet{Erkal:2019} and \citet{Shipp:2021}. We fix the scale radius to match the circular velocity measurement of $91.7$ km/s at $8.7$ kpc from \cite{vanderMarel:2014}. Note that this is in agreement with more recent measurements of the LMC's circular velocity \citep[e.g.,][]{Cullinane:2020}. We account for the dynamical friction of the Milky Way on the LMC using the results of \cite{Jethwa:2016}. We also fix the LMC's present-day proper motion, distance, and radial velocity to measured values \citep{Kallivayalil:2013,Pietrzyski:2013,van_der_marel:2002}. 
The LMC mass remains fixed throughout each simulation.

We model the potential of the Jet stream's progenitor as a Plummer sphere \citep{Plummer:1911} with a mass and scale radius chosen to match the observed stream width. During the course of tidal disruption, the progenitor's mass decreases linearly in time to account for tidal stripping. Since Jet does not have a known progenitor, we assume that the progenitor has completely disrupted, i.e., that its present day mass is zero. Furthermore, we assume that the remnant of the progenitor is located at $\phi_1=0\degree$.

We calculate the likelihood for the stream model by producing a mock observation of a simulated stream and comparing it with the data described in the previous sections. 
For each stream model, we calculate the track on the sky, the radial velocity, the proper motions in $\phi_1$ and $\phi_2$, and the distance as functions of $\phi_1$, the observed angle along the stream. 

We assign the mass of the progenitor in order to reproduce the observed width of the stream. 
Our best-fit model uses a progenitor mass of $M_{\rm prog}=2 \times 10^4 \Msun$, and a Plummer scale radius of $r_{\rm plum}=10 \pc$. 
We note that these values are highly dependent on the location of the progenitor. 

\begin{figure}[thb]
    \centering
    \includegraphics[width=0.98\columnwidth]{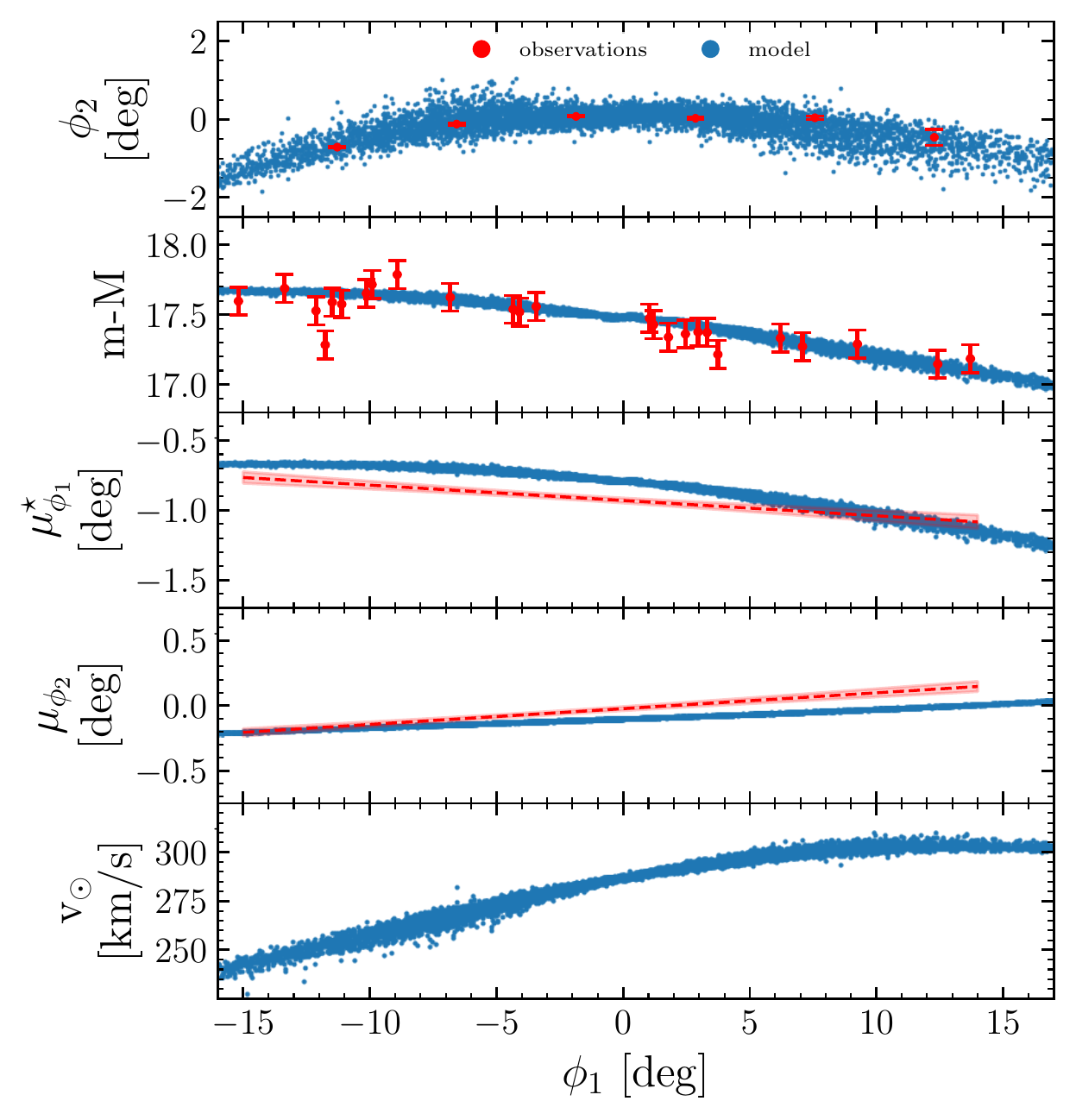}
    \caption{ Best-fit stream model to the Jet stream. In each row the dark blue points show the best-fit stream model and the red points show the observations to which the model was fit. \emph{First row:} the on-sky distribution of the stream. \emph{Second row:} the distance modulus of the stream. 
    \emph{Third and Fourth rows:} The reflex-corrected proper motions of the stream in $\phi_1$ and $\phi_2$ respectively. 
    \emph{Fifth row:} predicted  heliocentric velocities of the stream.
    }
    \label{fig:model}
\end{figure}

We perform a Markov Chain Monte Carlo (MCMC) fit using \code{emcee} \citep{Foreman_Mackey:2013}. Our model includes 5 free parameters. We fit the present-day progenitor $\phi_2$ position, distance, radial velocity, and proper motion.  
The prior distributions on each parameter are listed in \tabref{priors}.
The position of the progenitor along the stream is fixed to $\phi_1 = 0\degree$ (i.e., the middle of the stream's observed extent).
We show the Jet data and the best-fit stream models in  \figref{model}. 
In each panel we show the observations in red, and simulated stream in blue.
The radial velocity panel contains no observations, but can be used to predict the radial velocity of the stream. 
We find the best-fit model is a good fit to the observations of the distance modulus, stream track and proper motions.

We have tried fits that include/exclude the effect of the LMC and Milky Way bar. For the Milky Way bar we assume an analytic model with the same parameters as used in \citet[]{Li:2020} (described in their Section 5.2.1) and \citet{Shipp:2021}. For both cases, the LMC and Milky Way bar, we find that it is unlikely that the Jet stream has been significantly affected by these substructures.

This model emphasizes some of the observed features of the Jet stream discussed in Sections \ref{sec:morphology} and \ref{sec:proper_motion}. \response{In particular, we note the observed curvature of the stream track away from $\phi_2=0\degree$ is due both to Galactic parallax and the non-spherical potential of the MW.}
None of the intensity features gaps/peaks are seen in the model, and we also fail to replicate the off-track features seen in the photometry.
\response{Additionally, this model predicts a stream component in the Jet Bridge region (figure \ref{fig:search_region}) that is not detected in our search. 
This non-detection is likely due to the increased Milky Way foreground contamination and reddening in this region}

\begin{deluxetable*}{l ccccc}
\tablecolumns{13}
\tablewidth{0pt}
\tabletypesize{\scriptsize}
\tablecaption{ Priors on the dynamical model }
\label{tab:priors}
\tablehead{
\colhead{Parameter} & \colhead{Prior} & \colhead{Range} & \colhead{Units} & \colhead{Description} }
\startdata
$\phi_{\rm 2, prog}$ & Uniform & (-1, 1) & deg & Location of the progenitor perpendicular to the stream track. \\
$\mu_{\rm \alpha, prog}, \mu_{\rm \delta, prog}$ & Uniform  & (-10, 10) & mas/yr & Reflex-corrected proper motion of the progenitor. \\
$v_{\rm r, prog}$ & Uniform & (-500, 500) & km/s & Radial velocity of the progenitor. \\
$(m-M)_{\rm prog}$ & Normal  & $(m-M)_0 \pm 0.2 $ & mag & Distance modulus of the progenitor. \\
$\phi_{\rm 1, prog}$ & Fixed & 0 & deg & Location of the progenitor along the stream track. \\ 
\tableline
$M_{\rm LMC}$ & Fixed & $1.5 \times 10^{11}$ & \Msun & Total mass of the LMC. \\
$\mu_{\rm \alpha, LMC}$ & Normal  &  $ 1.91 $ & mas/yr & Proper motion of the LMC in RA. \\
$\mu_{\rm \delta, LMC}$ & Normal  & $ 0.229 $ & mas/yr & Proper motion of the LMC in Dec. \\
$v_{\rm r, LMC}$ & Normal  & $ 262.2 $ & km/s & Radial velocity of the LMC. \\
$d_{\rm LMC}$ & Normal  & $ 49970.0 $ & pc & Distance of the LMC.
\enddata
\end{deluxetable*}

\section{Discussion}
\label{sec:Discussion}

\subsection{Properties of the Jet stream}

The Jet stream is now detected from $-15.2 \degree \leq \phi_1 \leq 13.7 \degree$ increasing its known length from $\sim 11\degree$ to nearly $29 \degree$.
For $\sim 23 \degree$ ($-13 \degree \leq \phi_1 \leq 10 \degree$) the main sequence turn off of the stream is strongly detected in the \delve photometry; at $\phi_1 < -13 \degree$ and $> 10 \degree$ the intensity of the stream decreases greatly, and so is only significantly detected using BHB and RGB stars with measured proper motions.  
At observed distances ranging from $\sim 26-34.5 \kpc$, the stream has a physical length of 16 kpc, with a strong photometric detection covering 13.4 kpc.
This makes the extent of the Jet stream comparable to the kinematically cold Phoenix, ATLAS, and GD-1 streams \citep{Balbinot:2016, Koposov:2014,Grillmair:2006}, which span $\roughly 5 \kpc$ \citep[$13.6 \degree$;][]{Shipp:2018}, $\roughly 12 \kpc$ \citep[$34 \degree$ including Aliqa Uma;][]{Li:2020} and $\roughly 15.4 \kpc$ \citep[$\roughly 100 \degree$;][]{deBoer:2020,Webb:2019, Price-whelan:2018ApJ...863L..20P, Malhan:2018b}, respectively.
Our dynamical models place Jet on a retrograde orbit with a pericenter of \CHECK{$\rperi \kpc$} which is comparable to the pericenters of Phoenix \citep[$\roughly 13 \kpc$;][]{Wan:2020} ATLAS \citep[$ 13.3 \kpc$;][]{Li:2020}, 
\response{Kshir \citep[$\roughly 14 \kpc$;][]{Malhan:2019b}}, 
and GD-1 \citep[$\roughly 14 \kpc$;][]{koposov:2010ApJ...712..260K, Malhan:2019}. 

As discussed in Section \ref{sec:proper_motion}, the proper motion of the Jet stream members and the observed track of the stream are fairly well aligned, suggesting that the Jet stream has not been strongly perturbed perpendicular to the line of sight from interactions with large Milky Way substructures.  However, perturbations in the radial or track direction are difficult to measure from proper motion alone.

\subsection{Small-Scale Features}
\label{sec:discsmall}
This detailed view of the Jet stream has started to reveal its complexity, adding it to the group of streams that show small-scale features \citep[e.g.,][]{Erkal2017:pal5MNRAS.470...60E,Price-whelan:2018ApJ...863L..20P,Li:2020, Caldwell:2020}. 
Most noticeably, a $\roughly 4 \degree$ gap in the stream is seen centered on $\phi_1=-6 \degree$ extending from -. 
This structure may be due to interactions between the Jet stream and its environment (e.g., a dark matter subhalo passing by and perturbing the stream), or the environment of the progenitor parent satellite \citep[e.g.,][]{Malhan:2021}. 
Alternatively, this structure could also be the result of a complete dissolution of the progenitor as suggested by \citet{Webb:2019} in relation to the GD-1 stream. 
Understanding the nature of this gap will be important for future studies with deeper photometry and radial velocities of member stars near this region. 
We also note that in the top row of Figure \ref{fig:2dmodel} the stream looks extremely clumpy on smaller scales than our model probes. 
Deeper photometric observations, such as those possible with the Vera C.\ Rubin Observatory's Legacy Survey of Space and Time (LSST), will increase the signal-to-noise of these features, allowing better modeling and therefore a more complete understanding of the system. 
Finally, there are density features seen off the main track of the stream. At $\phi_1 \sim 5 \degree$ a signal is seen in the matched-filter map just above the stream (\figref{2dmodel}). 
This could be a substructure similar to the ``spur" of GD-1 \citep[]{Price-whelan:2018ApJ...863L..20P} or evidence of some other type of interaction. 

At ~$\phi_1 \sim -12.5\degree$, a feature is seen both above and below the stream; it is possible that the increased reddening in this region is causing this feature, or it could be even more evidence of past interactions between the Jet stream and other substructure. 

To fully understand these features, followup spectra will be crucial, as these observations will enable the use of radial velocities and metallicities to robustly identify members, as well as allow for the use of the 6D information of members in these off track features to determine their origin. 

The observations of small-scale structure in the Jet stream are particularly interesting given its orbital properties. 
The Jet stream's current Galactocentric radius ($r_G=30-37$ kpc), orbital pericenter ($r_{peri}=\rperi$ kpc), lack of perturbation from the Milky Way bar in our simulations, and retrograde orbit suggest the Jet stream is less likely to have been perturbed on small scales due to interaction with baryonic matter \citep{Pearson:2017,Banik:2021}. 
This indicates that  the Jet stream is likely to be one of the best known streams for constraining dark matter substructure in the Milky Way. 

\subsection{Stream Mass and Progenitor Properties}

Based on the \delve photometry, Jet appears to be another stellar stream with no obvious detected progenitor (e.g., Phoenix, ATLAS, GD-1, and many others). 
Although we do not detect a progenitor for Jet, we can use our observations and modeling to further constrain the properties of the progenitor of the Jet stream. 
J18 determined the current stellar mass of the Jet stream by fitting the observed stream-weighted CMD and found that the total stellar mass is  $2.5 \pm 0.2 \times 10^4 \Msun$.  
We can set a lower limit on the stellar mass of the Jet stream using the number of high confidence BHB candidates we detect. Based on the color--magnitude and proper motion selections applied in Sections \ref{sec:gradient} and \ref{sec:proper_motion}, we detect 28 high confidence BHB candidates along the stream. 
Assuming a \citet{Chabrier:2001} initial mass function, an age of 12.1\Gyr and a metallicity of [Fe/H]$=$\fehjet, we find that a stellar mass of $2.6_{-0.4}^{+0.6} \times 10^4 \Msun$ is required to produce the observed number of BHB stars. 
This estimate is in good agreement with the previous results of J18.

\citet{Erkal:2016} suggested that the total dynamical mass of a stream can be estimated from its width.
This was used by \citet{Shipp:2018} to estimate dynamical masses of the DES streams, and J18 applied the same procedure to estimate that the total dynamical mass of the Jet progenitor is expected to be $\roughly 6.4 \times 10^{4} \Msun$.
In this analysis, we find  the observed angular width of Jet varies by a factor of $\roughly 2$ over its extent (\secref{morphology}). However, if we account for the measured distance gradient (\secref{gradient}), the observations are consistent with a constant physical width of $\sim 90 \pc$ over the range $\phi_1=-12 \degree -10 \degree$. 
The stream appears to fan out even more in the region $\phi_1 > 10 \degree$ where the intensity of the stream drops greatly.
The observed complex physical structure makes it difficult to motivate the simple scaling between stream width and dynamical mass from \citet{Erkal:2016}.
Thus, we instead estimate the dynamical mass from the best-fit orbital model of the Jet stream described in \secref{modeling}.
We find that these simulations prefer a total dynamical mass of $\roughly 2 \times 10^{4} \Msun$ which is a factor of $\roughly3$ less massive than the estimate of J18, but our mass estimate is highly dependent on the location of the progenitor which we assume is $\phi_1=0 \degree$. 
For a progenitor at $\phi_1=20 \degree $ we find a worse fit to the overall stream properties but recover a dynamical mass consistent with J18.
The ratio of stellar and dynamical mass ($\roughly 1$) supports the hypothesis that the progenitor of Jet was a globular cluster \citep[$\mathrm{M/L} \, \roughly 1-2$;][]{Kruijssen:2008} rather than a dwarf galaxy \citep[$\mathrm{M/L}\, \roughly 10^2-10^4$ for $\mathrm{L}\, \roughly 2\times 10^4$;][]{Simon:2019}. 

The results of our MCMC modeling can be used to estimate the heliocentric velocity ($\mathrm{v}_\odot$) of the stream and other orbital parameters.  
We find a predicted heliocentric velocity at $\phi_1=0 \degree$ of \CHECK{$\mathrm{v}_{\odot}=286\pm10 \km/\second$}.\footnote{This agrees with unpublished spectroscopic data from AAT/2dF (T.~S.~Li, private communication).} For the orbit of Jet we find a pericenter of \CHECK{$r_{\rm peri}=\rperi\pm \rperierr \kpc$}, an apocenter of \CHECK{$r_{\rm apo}=\rapo \pm \rapoerr \kpc$}, and an eccentricity of \CHECK{0.59} and an orbital period of \CHECK{0.5 Gyr}. 

These orbital properties can be used to explore whether Jet could be associated with other known globular clusters or streams.
The predicted $R_v$ estimate, along with the measured proper motion of the Jet stream, give an expected angular momentum perpendicular to the Galactic disk ($\hat{z}$ direction) $L_z$ and total energy $E_{\rm tot}$. 
These two quantities, $L_z$ and $E_{\rm tot}$, are both conserved assuming a static axis-symmetric Milky Way potential. 
To compute these parameters we use the same Milky Way plus LMC potential from Section \ref{sec:modeling}.
We randomly draw from the posteriors of our fit values for the proper motions, $\phi_2$ position and radial velocity of the Jet stream at $\phi_1=0 \degree$, and repeat this 1000 times. Then for each draw we compute the $L_z$ and $E_{tot}$ of the Jet stream at $\phi_1=0\degree$. 
Doing this we find the predicted $L_z$ and $E_{\rm tot}$ of the Jet stream to be $L_z=\lz\pm\lzerr \kpc^2/\text{Myr}$ and predicted $E_{\rm tot}=\energy\pm\energyerr \kpc^2/\text{Myr}^2$. 
Using these results we look for globular clusters with similar $E_{\rm tot}$ and $L_z$ properties that could have been the progenitor of the Jet stream. 
From the \citet{Vasiliev:2019} catalog of globular cluster orbital properties we find no close matches suggesting that the progenitor of the Jet stream is either fully disrupted or undiscovered.

Comparing these results to Figures 1 and 2 in \citet{Bonaca:2020}, we find that the Jet stream is on a retrograde orbit with orbital parameters closest to Phelgethon ($L_z \roughly 1.93 \kpc^2/\Myr
$, $E_{\rm tot}\roughly-0.10 \kpc^2/\Myr^2$) and nearby Wambelong and Ylgr as well.
It seems likely that the progenitors of Phelgethon and Jet were accreted onto the Milky Way in the same accretion event. 
The work of \citet{Naidu:2020} with the H3 survey identified a number of Milky Way accretion events, and localized them in the $E_{tot}-L_z$ paramter space. 
The $E_{tot}-L_z$ properties of the Jet stream places it in the region of parameter space likely associated with the Sequoia, I'itoi, and Arjuna progenitors (their Figure 2), suggesting that the progenitor of the Jet stream was a globular cluster associated with one of these accretion events \citep{Naidu:2020,Bonaca:2020, Myeong:2019}.  

\response{Similarly to J18, we note that the stellar stream PS1-B \citep{Bernard:2016} is well aligned to the on-sky track of the Jet stream, but at a much different distance ($D_\odot=14.5 \kpc$). In fact, with our extended detection to large $\phi_1$ with BHB stars the two on-sky stream tracks come within $0.3\degree$ of each other.
Our matched filter analysis did not detect the PS1-B stream; however, our filter was not optimized for the detection of PS1-B, and future studies could further investigate this potential association.}

\section{Conclusions}
\label{sec:conclusions}

We have presented deep photometric and astrometric measurement of the Jet stream. 
We utilized the deep, wide-field, homogeneous DELVE DR1 data, which allowed us to discover substantial extensions of the Jet stream.
We used both \delve photometry and proper motions from \Gaia EDR3 to select a sample of candidate BHB member stars. 
These stars allow us to resolve a distance gradient along the stream.
The \delve photometry is then used to model the stream intensity, track, and width, quantitatively characterizing the observed density variations. 
Additionally, we are able to use BHB and RGB stars to measure the systemic proper motion and proper motion gradient of Jet for the first time.
Finally, we fit the stream with a dynamical model to constrain the orbit of the Jet stream. 

The results of these analyses are summarized as follows:
\begin{itemize}
    \item We extend the known extent of the Jet stream from $11 \degree$ to $\roughly29\degree$ corresponding to a physical length of $\roughly16$ kpc.
    \item We measure a distance gradient of $-0.2$ kpc/deg along the stream ranging from $D_\odot \sim 34.2 \kpc$ at $\phi_1=-15 \degree$ to
    $D_\odot \sim 27.4 \kpc$ at $\phi_1=13.7\degree$. 
    \item We model the stream morphology to quantitatively characterize the stream track, width and linear density. We identify a gap in the stream and two features off the main track of the stream. 
    \item We measure the proper motion of the Jet stream for the first time, and identify likely member RGB/BHB stars from their proper motions. 
    \item Our modeling suggests Jet is on a retrograde orbit, unlikely to have been significantly affected by the LMC or Milky Way bar, and has an orbital pericenter of $r_{peri}=\rperi$ kpc.
\end{itemize}

Our analysis of the Jet stream has already been used to target spectroscopic measurements with AAT/2dF as part of the Southern Spectroscopic Stellar Stream Survey \citep[\SSSSS;][]{Li:2018}.
Medium-resolution spectroscopic measurements with \SSSSS will confirm stream membership, provide radial velocities for stream members, and measure metallicities from the equivalent widths of the calcium triplet lines \citep{Li:2018}. 
Such measurements have already yielded interesting dynamical information for the ATLAS stream \citep{Li:2020} and measured an extremely low metallicity for the Phoenix stream \citep{Wan:2020}. 
These measurements will further allow the targeting of high-resolution spectroscopy, which can provide detailed elemental abundances for Jet member stars \citep{Ji:2020}, and help to determine the nature of the Jet stream progenitor. 

The future of resolved stellar studies is bright with ongoing and future deep and wide-area photometric surveys. In particular, detailed studies of stellar streams will provide important information for modeling both the large and small-scale structure of the Milky Way halo, ultimately helping to constrain the fundamental nature of dark matter \citep{Drlica-Wagner:2019}. In the near future, \delve will significantly improve the extent and homogeneity of the southern sky coverage, setting the stage for the LSST-era. Our work on the Jet stream provides an important precursor legacy to similar measurements that will be possible with LSST.

 \section{Acknowledgments} 
PSF acknowledges support from from the Visiting Scholars Award Program of the Universities Research Association. 
ABP acknowledges support from NSF grant AST-1813881.
The DELVE project is partially supported by Fermilab LDRD project L2019-011 and the NASA Fermi Guest Investigator Program Cycle 9 No. 91201.

This project used data obtained with the Dark Energy Camera (DECam), which was constructed by the Dark Energy Survey (DES) collaboration.
Funding for the DES Projects has been provided by 
the DOE and NSF (USA),   
MISE (Spain),   
STFC (UK), 
HEFCE (UK), 
NCSA (UIUC), 
KICP (U. Chicago), 
CCAPP (Ohio State), 
MIFPA (Texas A\&M University),  
CNPQ, 
FAPERJ, 
FINEP (Brazil), 
MINECO (Spain), 
DFG (Germany), 
and the collaborating institutions in the Dark Energy Survey, which are
Argonne Lab, 
UC Santa Cruz, 
University of Cambridge, 
CIEMAT-Madrid, 
University of Chicago, 
University College London, 
DES-Brazil Consortium, 
University of Edinburgh, 
ETH Z{\"u}rich, 
Fermilab, 
University of Illinois, 
ICE (IEEC-CSIC), 
IFAE Barcelona, 
Lawrence Berkeley Lab, 
LMU M{\"u}nchen, and the associated Excellence Cluster Universe, 
University of Michigan, 
NSF's National Optical-Infrared Astronomy Research Laboratory, 
University of Nottingham, 
Ohio State University, 
OzDES Membership Consortium
University of Pennsylvania, 
University of Portsmouth, 
SLAC National Lab, 
Stanford University, 
University of Sussex, 
and Texas A\&M University.

This work has made use of data from the European Space Agency (ESA) mission {\it Gaia} (\url{https://www.cosmos.esa.int/gaia}), processed by the {\it Gaia} Data Processing and Analysis Consortium (DPAC, \url{https://www.cosmos.esa.int/web/gaia/dpac/consortium}).
Funding for the DPAC has been provided by national institutions, in particular the institutions participating in the {\it Gaia} Multilateral Agreement.

This work is based on observations at Cerro Tololo Inter-American Observatory, NSF's National Optical-Infrared Astronomy Research Laboratory (2019A-0305; PI: Drlica-Wagner), which is operated by the Association of Universities for Research in Astronomy (AURA) under a cooperative agreement with the National Science Foundation.

This manuscript has been authored by Fermi Research Alliance, LLC, under contract No.\ DE-AC02-07CH11359 with the US Department of Energy, Office of Science, Office of High Energy Physics. The United States Government retains and the publisher, by accepting the article for publication, acknowledges that the United States Government retains a non-exclusive, paid-up, irrevocable, worldwide license to publish or reproduce the published form of this manuscript, or allow others to do so, for United States Government purposes.

\facility{Blanco, \Gaia.}
\software{\code{astropy} \citep{astropy:2013,astropy:2018}, \emcee \citep{Foreman_Mackey:2013}, \code{fitsio},\footnote{\url{https://github.com/esheldon/fitsio}} 
\healpix \citep{Gorski:2005},\footnote{\url{http://healpix.sourceforge.net}} \code{healpy},\footnote{\url{https://github.com/healpy/healpy}} \code{Matplotlib} \citep{Hunter:2007}, \code{numpy} \citep{numpy:2011},
\code{pandas}\citep{pandas:2020,mckinney-proc-scipy-2010},
\code{scipy} \citep{scipy:2001}, 
\code{STAN}\citep{Carpenter:2017}
\ugali \citep{Bechtol:2015}.\footnote{\url{https://github.com/DarkEnergySurvey/ugali}}}

\bibliographystyle{aasjournal}
\bibliography{main}

\appendix
\numberwithin{table}{section}
\section{Membership Probability}\label{sec:memprob}
Table \ref{tab:memprob} includes probable stream member stars with
membership probability greater than 0.1 from the likelihood analysis described in Section \ref{sec:proper_motion}.

\begin{table}[!ht]
    \small
    \hskip-2.2cm\begin{tabular}{c c c c c c c c c c}
    \hline
    \hline 
    \\
    ID (DELVE)\tablenotemark{a} & ID (Gaia)\tablenotemark{a} &  R.A.\tablenotemark{b} & Dec.\tablenotemark{b}  & $g_0$\tablenotemark{c} & $r_0$\tablenotemark{c} & $\mu_{\alpha} \cos{\delta}$  &  $\mu_{\delta}$ & $D_{\phi_1}$\tablenotemark{d} & $p_i$\tablenotemark{e} \\ 
     & & (deg) & (deg) & (mag) & (mag) & (mas/yr) & (mas/yr) & (kpc)& \\
    \hline
    
        10728400085912 & 5692169948945471488 & 147.32085 & -10.90436 & 18.06 & 17.51 & $-0.81 \pm 0.11$ & $-1.85 \pm 0.08$ & 27.28 & $0.83 \pm 0.23$ \\
        10728500064937 & 3769827073557279232 & 148.25261 & -10.92857 & 18.84 & 18.34 & $-0.93 \pm 0.21$ & $-1.29 \pm 0.28$ & 27.17 & $0.48 \pm 0.17$ \\
        10728500080043 & 3769884424255747584 & 147.84902 & -10.82876 & 18.25 & 17.71 & $-1.25 \pm 0.12$ & $-1.56 \pm 0.11$ & 27.20 & $0.38 \pm 0.22$ \\
        10728500026630 & 3769900397239081472 & 148.34539 & -10.75644 & 19.10 & 18.65 & $-0.92 \pm 0.21$ & $-2.20 \pm 0.24$ & 27.13 & $0.99 \pm 0.01$ \\
        10728500315132 & 3769918715274616320 & 148.13991 & -10.57946 & 19.01 & 18.47 & $-1.43 \pm 0.25$ & $-1.43 \pm 0.30$ & 27.12 & $0.87 \pm 0.07$ \\
        10728500024967 & 3770009145810545664 & 148.73686 & -10.33853 & 18.97 & 18.47 & $-1.13 \pm 0.23$ & $-2.04 \pm 0.23$ & 27.01 & $0.99 \pm 0.00$ \\
        10754000190703 & 5690650591379140608 & 145.22603 & -13.49643 & 18.82 & 18.35 & $-1.38 \pm 0.17$ & $-1.93 \pm 0.15$ & 27.99 & $0.74 \pm 0.17$ \\
        10754000071117 & 5690723434024204288 & 146.34768 & -13.10071 & 19.10 & 18.66 & $-0.96 \pm 0.20$ & $-1.55 \pm 0.17$ & 27.78 & $0.94 \pm 0.04$ \\
        10754000135902 & 5690754357789028352 & 146.29135 & -12.69526 & 18.68 & 18.15 & $-0.61 \pm 0.16$ & $-1.65 \pm 0.13$ & 27.72 & $0.90 \pm 0.13$ \\
        10754000148096 & 5690771365860339712 & 146.78020 & -12.57939 & 19.16 & 18.70 & $-1.63 \pm 0.23$ & $-1.56 \pm 0.21$ & 27.64 & $0.32 \pm 0.14$ \\

    \ldots & \ldots & \ldots & \ldots& \ldots& \ldots& \ldots& \ldots& \ldots& \ldots\\
    \end{tabular}
    \caption{this table includes all stars with $p_i > 0.1$ for our analysis of the proper motion of the Jet stream.\\
    (This table is available in its entirety in machine-readable form.)\\
    (a) DELVE ID's are from the \code{QUICK\_OBJECT\_ID} column in DELVE-DR1, and Gaia IDs are from \code{SOURCE\_ID} column in \gaia EDR3. \\
    (b) R.A. and Dec. are from Gaia EDR3 catalog (J2015.5 Epoch).\\
    (c) $g$, $r$ band magnitudes are reddening corrected PSF photometry (\code{MAG\_PSF\_DERED}) from DELVE DR1 catalog.\\
    (d) The $D_{\phi_1}$ column gives the distance in kpc derived from equation \ref{eqn:grad}, except for candidate bhb stars whose distances are estimated from their predicted absolute magnitude $M_g$ as discussed in section \ref{sec:gradient}}
    (e) The probability that a star is a member of the Jet stream. \\
    \label{tab:memprob}
\end{table}

\end{document}